\documentclass[twocolumn,aps,prl,superscriptaddress,nobalancelastpage,floatfix]{revtex4-2}

\usepackage[T1]{fontenc}
\usepackage[latin9]{inputenc}
\usepackage{lmodern}

\usepackage{amsmath}
\usepackage{amssymb}
\usepackage{graphicx}
\usepackage{bbm}	
\usepackage{tikz}
\usepackage{tikz-cd}

\usepackage{etoolbox}

\usepackage[colorlinks=true,citecolor=blue,linkcolor=red,breaklinks]{hyperref}

\definecolor{URLColor}{HTML}{0B6E4F}   

\makeatletter
\@ifundefined{textcolor}{}{
 \definecolor{BLACK}{gray}{0}
 \definecolor{WHITE}{gray}{1}
 \definecolor{RED}{rgb}{1,0,0}
 \definecolor{GREEN}{rgb}{0,1,0}
 \definecolor{BLUE}{rgb}{0,0,1}
 \definecolor{CYAN}{cmyk}{1,0,0,0}
 \definecolor{MAGENTA}{cmyk}{0,1,0,0}
 \definecolor{YELLOW}{cmyk}{0,0,1,0}
}
\makeatother

\usepackage{tikz-cd}
\usepackage{tikz}
\usetikzlibrary{decorations.pathreplacing,calligraphy,decorations.markings,calc}

\definecolor{tensor}{rgb}{1,1,1}
\definecolor{isometry}{rgb}{1,1,1}
\definecolor{invertibleM}{rgb}{1,1,1}
\definecolor{vector}{rgb}{1,1,1}
\definecolor{unitary}{rgb}{0.8,0.8,1}
\definecolor{gate}{rgb}{1.0,1.0,1.0}

\newcommand{\ATensorMPO}[2]{
	\begin{scope}[shift={(#1)}]
		\draw (-1.2,0) -- (1.2,0);
		\draw (0,1.2) -- (0,0);
        \draw (0,-1.2) -- (0,0);
		\filldraw[fill=tensor] (-0.6,-0.6) -- (-0.6,0.6) -- (0.6,0.6) -- (0.6,-0.6) -- (-0.6,-0.6);
		\draw (0,0) node {\scriptsize #2};
	\end{scope}
}

\newcommand{\xTensor}[2]{
	\begin{scope}[shift={(#1)}]
		\draw (-1,0) --(1,0);	   
        \filldraw[fill=invertibleM] (0,0) circle[radius=0.5]; 
    \node at (0,0) {\scriptsize #2}; 
	\end{scope}
}

\newcommand{\wTensor}[2]{
	\begin{scope}[shift={(#1)}]
		\draw (0,-0.3) --(0,1.3);	   \filldraw[fill=isometry,shift={(0,0)}](0,0.5) circle (0.5);
		\draw (0,0.5) node {\scriptsize #2};
	\end{scope}
}

\definecolor{ZM}{rgb}{.7,0.3,0}
\definecolor{SD}{rgb}{0,1,0}

\makeatother

\usepackage{babel}
\begin{document}
\title{Anomaly and symmetry-charge flow in mixed states }
\author{Ze-Min Huang }


\author{Sebastian Diehl}

\affiliation{Institute for Theoretical Physics, University of Cologne, 50937 Cologne, Germany}

\begin{abstract}
The $(1+1)$-dimensional chiral anomaly is a paradigmatic exact result in quantum field theory, traditionally formulated for zero-temperature pure states where it arises from spectral flow induced by external gauge fields and captures universal ground-state properties. In mixed states, however, the participation of many states and charge exchange with the environment invalidate this mechanism. Naive extensions yield model-dependent anomaly coefficients, calling its universality into question.
Here, we resolve this problem for Abelian symmetries by deriving the anomaly from an algebraic relation between the symmetry and its flux-insertion operator. We obtain symmetry-charge flow, a mixed-state generalization of spectral flow, in which an applied field redistributes statistical weight across symmetry-resolved charge sectors. Fixed solely by symmetry, the anomaly restores universality and applies to both pure and mixed states in fermionic and bosonic systems.
We substantiate these results in tight-binding fermionic models with continuous symmetry and in spin models with discrete symmetries.
\end{abstract}

\maketitle

\textit{\color{red}{Introduction.--}}
Many-body systems are inherently complex. The large number of degrees of freedom renders exact solutions exceptional, and genuinely non-perturbative principles are rare. Among the few are 't Hooft anomalies~\cite{hooft1980springer}, obstructions to
consistently gauging global symmetries. Such anomalies impose exact,
symmetry-based constraints on collective behavior. In
$\left(1+1\right)$ dimensions, for instance, they exclude a unique
gapped symmetric ground state~\cite{chen2011prb_czx}.
A paradigmatic case is the chiral anomaly~\cite{treiman1985princeton,bertlmann2000oxford,fujikawa2004oxford}.
Its current algebra forms the backbone of bosonization~\cite{treiman1985princeton,francesco2012springer,fradkin2013cambridge},
a cornerstone of our understanding of strongly correlated systems in low dimensions~\cite{giamarchi2004clarendon}. It enables the exact solution of the Schwinger model~\cite{roskies1981prd}, fixes fractional soliton charge~\cite{goldstone1981prl}, and underlies bulk-boundary correspondence in topological phases~\cite{callan1985npb,stone1991aop,qi2011rmp,shinsei2012prb1}.

\begin{figure}
\includegraphics[scale=0.18]{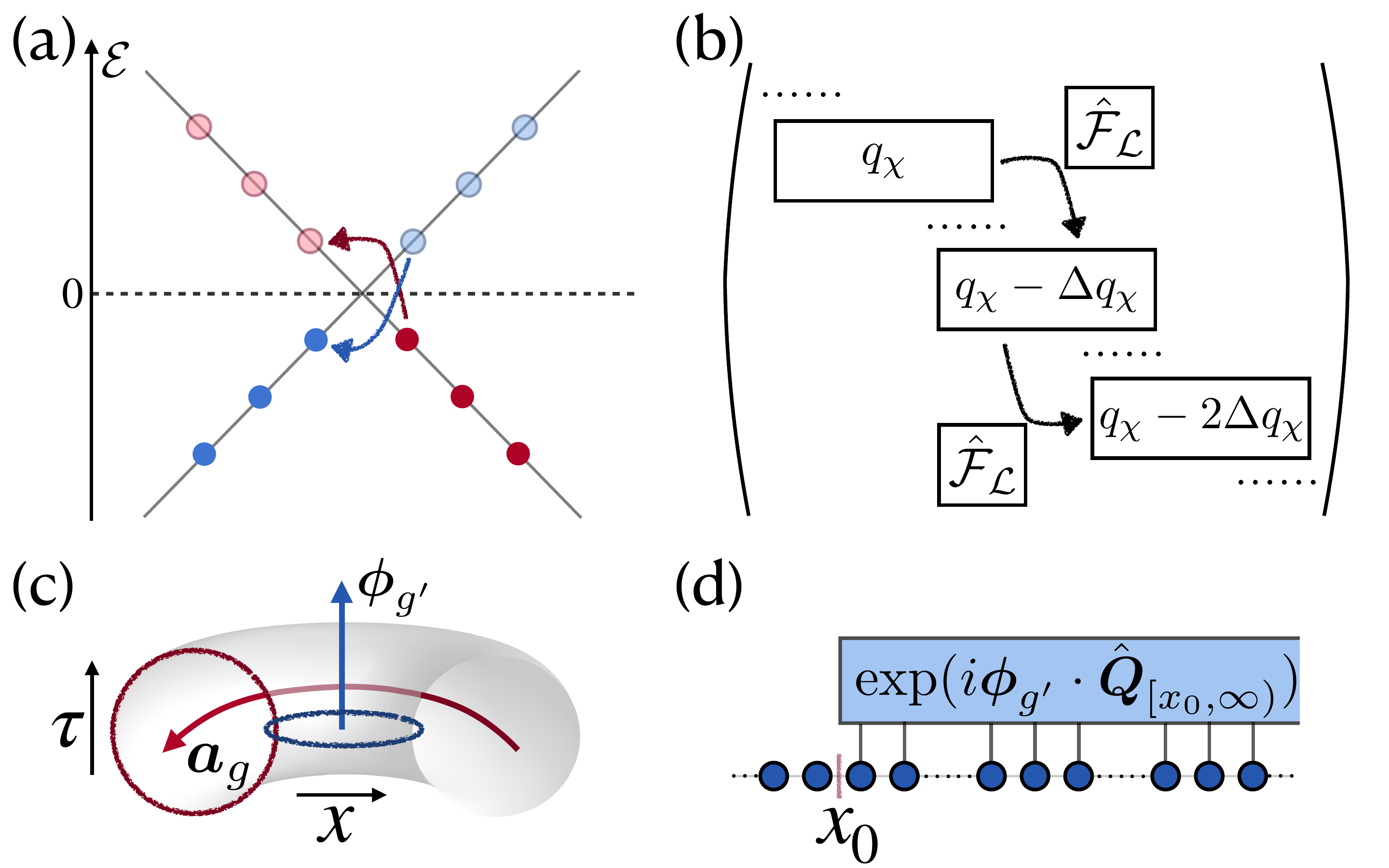}\caption{Conceptual illustration of spectral flow (a) and symmetry-charge flow (b) induced by a large gauge transformation; (c) temporal (red, $\boldsymbol{a}_g$) and spatial (blue, $\boldsymbol{\phi}_{g^\prime}$) fluxes; and (d) the spatial flux-insertion operator (blue, $\boldsymbol{\phi}_{g^\prime}$). For a Dirac fermion (a), a large gauge transformation (i.e., $\oint A_x\rightarrow \oint A_x+2\pi$) generated by the operator $\hat{\mathcal{F}}_{\mathcal{L}}$, creates a left-handed mode (red circle) from the ground state and annihilates a right-handed one (blue), reducing the net chiral charge by $2$ and moving the ground state to a different charge sector. Going beyond this ground-state picture, the anomaly persists for mixed states (b), i.e., a large gauge transformation permutes symmetry-charge sectors of the density matrix. (c, d) Temporal and spatial fluxes implement symmetry twists for particles winding around the temporal/spatial cycles. In particular, the spatial flux-insertion operator can be constructed by applying the symmetry transformation on half of an infinite system [see (d)].\label{fig:Illustration}}
\end{figure}

The chiral anomaly arises for Weyl fermions of opposite chirality and captures long-distance features of the ground state. Classically independent
$U(1)$ symmetries for right- and left-handed fermions are broken by quantum
fluctuations, encoded in the non-commutativity of the local symmetry
generators~\cite{treiman1985princeton},
\begin{equation}\label{eq:current_algebra}
\left[\hat{j}^{0}(x),\ \hat{j}_{\chi}^{0}(y)\right]
=-i\frac{1}{\pi}\partial_{x}\delta\left(x-y\right).
\end{equation}
Here, $\hat{j}^0(x)$ denotes the total charge density and
$\hat{j}_\chi^0(x)$ the chiral charge density. Equation~\eqref{eq:current_algebra}
is scale invariant~\cite{francesco2012springer} and can be derived
entirely from ground-state properties (see, e.g.,~\cite{fradkin2013cambridge});
it therefore constitutes an emergent identity for zero-temperature critical states. Physically, it expresses spectral flow~\cite{nielsen1983plb,fradkin2021princeton,arouca2022sp}: in a background $U(1)$ gauge field $A_\mu$, it is equivalently written
as the non-conservation of the chiral current,
\begin{equation}
\partial_{\mu}\langle\hat{j}_{\chi}^{\mu}\rangle_{G}
=-\frac{1}{\pi}\epsilon^{\mu\nu}\partial_{\mu}A_{\nu},
\label{eq:chiral_anomaly}
\end{equation}
with $\langle\dots\rangle_{G}$ the ground-state expectation value.
Inserting one flux quantum enforces level crossings with the ground
state [Fig.~\ref{fig:Illustration}(a)], changing the net chiral charge
by $-2$, as required by Eq.~\eqref{eq:chiral_anomaly}.

This formulation, however, relies on an idealized zero-temperature
pure state. Realistic quantum systems are generically in mixed states,
whether due to finite temperature or system-environment coupling. Extending
't Hooft anomalies to mixed states is therefore both necessary and
non-trivial. Two obstacles arise immediately:
(i) mixed states involve contributions from many states,
whereas Eq.~\eqref{eq:current_algebra} is governed by ground-state properties; and
(ii) charge exchange with an environment invalidates current
conservation as a sharp diagnostic of the anomaly.
Accordingly, recent classification schemes for open systems typically
assume that at least one symmetry charge remains conserved
\cite{kawabata2024prl,lessa2025prx,wang2025prxq,sun2025arxiv}; the corresponding classification can become trivial otherwise~\cite{wang2025prxq}. For systems permitting charge exchange,
diagnostics have been developed for exemplary cases
\cite{huang2022prb,zhou2024nsr,huang2025prl,zang2024prl,
huang2025prl_mspt,mao2025cpl,xu2025arxiv,ma2025arxiv,shi2025arxiv, sala2026qst},
but a general and symmetry-based resolution is still missing.

In this Letter, we fill this gap, by providing a symmetry-based construction for mixed-state anomalies in $(1+1)$ dimensions with Abelian symmetries, both continuous and discrete. The central step is to derive the flux-insertion operator solely from the symmetry operator itself. This renders the construction independent of microscopic details and state purity, and explains its universality across many-body systems. The resulting algebra generalizes
Eq.~\eqref{eq:current_algebra} to mixed states by acting on the full Hilbert space, thereby resolving (i). 
Moreover, it captures charge redistribution under large gauge transformations [Fig.~\ref{fig:Illustration}(b)], which we term \emph{symmetry-charge flow}. In the zero-temperature limit, this reduces to spectral flow, manifested as the charge creation or
annihilation relative to the ground state encoded in
Eq.~\eqref{eq:chiral_anomaly}. For mixed states, however, the redistribution extends across the full Hilbert space, providing a state-independent diagnostic that does not rely on a continuity
equation and thus overcomes (ii). 
We demonstrate our construction in fermionic models with continuous
symmetries and in spin systems with discrete symmetries.

\textit{\color{red}{Coupling partition functions to symmetry fluxes.--}}
Consider a $G$-symmetric density matrix $\hat\rho$,
\begin{equation}\label{eq:sym_rho}
\hat U_g \hat\rho \hat U_g^\dagger=\hat\rho,
\qquad
\hat U_g \equiv e^{i\boldsymbol a_g\cdot \hat{\boldsymbol Q}},
\qquad
\forall\, g\in G,
\end{equation}
a condition commonly referred to as weak symmetry.
Here $G$ is Abelian, $G=\prod_I \mathbb Z_{\mathcal N_I}$, with commuting charge operators $\hat Q_I$ satisfying $e^{i2\pi \hat Q_I}=\mathbb I$.
Writing $\boldsymbol a_g\cdot \hat{\boldsymbol Q}\equiv \sum_I a_{g,I}\hat Q_I$, we take
$a_{g,I}\in \frac{2\pi}{\mathcal N_I}\{0,1,\dots,\mathcal N_I-1\}$.
Let $\mathcal N_g$ be the order of $g$, i.e., the smallest positive integer such that $(\hat U_g)^{\mathcal N_g}=\mathbb I$.
We then define the associated (integer-valued) charge operator
\begin{equation}\label{eq:symm_charge}
\hat Q_g \equiv \frac{\mathcal N_g}{2\pi}\,\boldsymbol a_g\cdot \hat{\boldsymbol Q},
\end{equation}
which reduces to the usual charge for continuous symmetries as $\mathcal N_g\to\infty$.

We couple $\hat\rho$ to background gauge fields, and characterize mixed-state anomalies by the resulting failure of gauge invariance of the partition function. Concretely, we implement the backgrounds as symmetry fluxes, so that discrete and continuous symmetries are treated on the same footing. The corresponding flux-inserted partition function is then \cite{FNtrace},
\begin{equation}\label{eq:partition_function_flux}
Z[\boldsymbol a_g,\boldsymbol\phi_{g'}]\equiv \mathrm{Tr}\!\left[\hat\rho(\boldsymbol\phi_{g'})\,e^{i\boldsymbol a_g\cdot \hat{\boldsymbol Q}}\right],
\qquad
g,g'\in G .
\end{equation}
Here $\boldsymbol\phi_{g'}$ is the spatial $g'$ flux [blue arrow in Fig.~\ref{fig:Illustration}(c)], so that traversing the spatial cycle once acts as $e^{i\boldsymbol\phi_{g'}\cdot\hat{\boldsymbol Q}}$, with
$\phi_{g',I}\in \frac{2\pi}{\mathcal N_I}\{0,1,\dots,\mathcal N_I-1\}$.
The corresponding flux-threaded density matrix is generated by a flux-insertion operator $\hat{\mathcal F}(\boldsymbol\phi_{g'})$ (constructed below for an infinite chain),
$\hat\rho(\boldsymbol\phi_{g'})=\hat{\mathcal F}(\boldsymbol\phi_{g'})\,\hat\rho(0)\,\hat{\mathcal F}(\boldsymbol\phi_{g'})^{-1}$.
Likewise, $\boldsymbol a_g$ denotes the temporal flux [red arrow in Fig.~\ref{fig:Illustration}(c)], such that $e^{i\boldsymbol a_g\cdot\hat{\boldsymbol Q}}$ is a symmetry of $\hat\rho$ and implements a global twist on a fixed time slice.
The definition \eqref{eq:partition_function_flux} does not rely on equilibrium: in the $T\to 0$ equilibrium limit it reduces to the Euclidean vacuum amplitude \cite{bertlmann2000oxford}, while out of equilibrium it yields the Keldysh partition function \cite{sieberer2016rpp}.
The flux-inserted partition function provides a unified anomaly diagnostic through its transformation under a large gauge transformation of the spatial flux,
\begin{equation}\label{eq:Z_thooft}
\boldsymbol\phi_{g'}\rightarrow \boldsymbol\phi_{g'}^{(\mathcal L)}:
\qquad
Z\!\left[\boldsymbol a_g,\boldsymbol\phi_{g'}^{(\mathcal L)}\right]
=
e^{i\Delta\mathcal S_g}\,
Z\!\left[\boldsymbol a_g,\boldsymbol\phi_{g'}\right],
\end{equation}
where $\Delta\mathcal S_g$ is a model-dependent phase (given explicitly below for the models of interest).
A non-trivial $\Delta{\mathcal{S}}_g$ signals a failure of gauge invariance, i.e., a 't-Hooft anomaly.
Physically, it is the braiding phase between temporal and spatial fluxes, as expected from anomaly inflow \cite{callan1985npb}.
Finally, while large gauge transformations of spatial fluxes need not detect all anomalies, they provide a useful diagnostic for a broad class, with the chiral anomaly as a special case (see below).

\textit{\color{red}{Mixed-state anomalies and symmetry-charge flow.--}}The phase in Eq.~\eqref{eq:Z_thooft} is insensitive to microscopic details. This can be made explicit by deriving a sufficient operator-algebraic condition involving only symmetry operators and their flux insertions. We therefore construct (i) a flux-insertion operator and (ii) the corresponding large-gauge-transformation operator, both deduced from the symmetry  [Eq.~\eqref{eq:sym_rho}] rather than taken as independent inputs.
For an infinite chain (for simplicity), a spatial $g'$ flux is inserted by applying the $g'$ transformation only to the half-infinite region to the right of a cut at $x_0$ \cite{zaletel2013prl,zaletel2014njp,huang2025prl_mspt},
\begin{equation}\label{eq:flux_insertion}
\hat{\mathcal F}(\boldsymbol\phi_{g'})\equiv e^{i\boldsymbol\phi_{g'}\cdot \hat{\boldsymbol Q}_{[x_0,\infty)}}.
\end{equation}
Crossing the cut [Fig.~\ref{fig:Illustration}(d)], a $g'$-charged excitation acquires the expected Aharonov--Bohm phase linear in $\boldsymbol\phi_{g'}$.
A large gauge transformation $\hat{\mathcal F}_{\mathcal L}$ relates equivalent representatives of the same background flux,
\begin{equation}\label{eq:large_gauge_F}
\hat{\mathcal F}(\boldsymbol\phi_{g'}^{(\mathcal L)})=\hat{\mathcal F}_{\mathcal L}\,\hat{\mathcal F}(\boldsymbol\phi_{g'}) ,
\end{equation}
and thus encodes the ambiguity in choosing a flux configuration for a given group element \cite{FN_sm}.
Namely, while $\hat U_{g_1}\hat U_{g_2}=\hat U_{g_1g_2}$, flux insertions compose only up to a large gauge transformation,
$\hat{\mathcal F}(\boldsymbol\phi_{g_1})\,\hat{\mathcal F}(\boldsymbol\phi_{g_2})
=
\hat{\mathcal F}_{\mathcal L}\,\hat{\mathcal F}(\boldsymbol\phi_{g_1g_2}).$
This implies that $\hat{\mathcal F}_{\mathcal L}$ is localized near the cut $x_0$, with distant contributions canceling; for continuous Abelian symmetries it relates fluxes differing by an integer flux quantum. 
The locality and invertibility of $\hat{\mathcal F}_{\mathcal L}$ are essential for the anomaly result below. On a finite periodic chain (see Appendix~\ref{supp_sec:flux_insertion} for details), the corresponding large-gauge transformation preserves these properties. By contrast, a generic flux insertion on a finite periodic chain need not be representable as a similarity transformation. For notational simplicity, we nevertheless retain the operator notation
$\hat{\mathcal F}(\boldsymbol{\phi}_{g^\prime})$
in the main text for the infinite-chain construction. The anomaly relations remain valid on finite periodic chains, as demonstrated in the Supplemental Material.

The operator-algebraic condition then follows from the two previously mentioned properties, (i) $e^{i\boldsymbol a_g\cdot\hat{\boldsymbol Q}}$ remains a symmetry before and after the gauge transformation, and (ii) locality of $\hat{\mathcal F}_{\mathcal L}$, which acts on the flux-threaded density matrix by conjugation,
$\hat\rho(\boldsymbol\phi_{g'}^{(\mathcal L)})=
\hat{\mathcal F}_{\mathcal L}\,\hat\rho(\boldsymbol\phi_{g'})\,\hat{\mathcal F}_{\mathcal L}^{-1}.$
Property (i) implies that
$[\hat{\mathcal F}_{\mathcal L}^{-1}e^{i\boldsymbol a_g\cdot\hat{\boldsymbol Q}}\hat{\mathcal F}_{\mathcal L},\,\hat\rho]=0$.
By locality, $\hat{\mathcal F}_{\mathcal L}$ cannot change the global symmetry operator $e^{i\boldsymbol a_g\cdot\hat{\boldsymbol Q}}$ except possibly by an overall phase [assuming $\hat{\rho}$ commutes only with the symmetry operators in Eq.~\eqref{eq:sym_rho}]. Thus, for $|Z|\neq 0$ one obtains the exact operator identity \cite{FNmath},
\begin{equation}\label{eq:flux_charge}
\hat{\mathcal F}_{\mathcal L}^{-1}\,e^{i\boldsymbol a_g\cdot\hat{\boldsymbol Q}}\,\hat{\mathcal F}_{\mathcal L}
=
e^{i\Delta\mathcal S_g}\,e^{i\boldsymbol a_g\cdot\hat{\boldsymbol Q}} .
\end{equation}
This equation is a mixed-state analogue of the current algebra \eqref{eq:current_algebra}, but holds as an exact statement on the full (finite) Hilbert space rather than only in the infrared. Finally, $|Z|=0$ for the maximally mixed (infinite-temperature) state whenever $\Delta\mathcal S_g\neq 0$.

Equation~\eqref{eq:flux_charge} admits a physical interpretation as a mixed-state generalization of spectral flow, which we refer to as \emph{symmetry-charge flow} in Hilbert space. Using \eqref{eq:sym_rho}, decompose $\hat\rho$ into $g$-charge sectors,
\begin{equation}\label{eq:rho_sector_decomp}
\hat\rho=\sum_{q_g} p_{q_g}\,\hat\rho_{q_g},
\quad
\hat\rho_{q_g}\equiv
\frac{\hat P_{q_g}\hat\rho\,\hat P_{q_g}}{\mathrm{Tr}(\hat\rho \hat P_{q_g})},
\quad
p_{q_g}\equiv \mathrm{Tr}(\hat\rho \hat P_{q_g}),
\end{equation}
where $q_g\in\mathbb Z$ are eigenvalues of $\hat Q_g$ in \eqref{eq:symm_charge}.
The projector onto fixed $g$-charge is
\begin{equation}\label{eq:P_g}
\hat P_{q_g}
=\delta(\hat Q_g-q_g)
=
\frac{1}{\mathcal N_g}
\sum_{x_g\in\frac{2\pi}{\mathcal N_g}\{0,1,\dots,\mathcal N_g-1\}}
e^{ix_g(\hat Q_g-q_g)},
\end{equation}
where the second equality is from the discrete Fourier representation. This representation is standard in full-counting statistics and is both numerically and experimentally accessible (see e.g., \cite{gross2017science,humeniuk2017prl, huang2025prr}).
A large gauge transformation then shifts the sector weights,
\begin{equation}\label{eq:symm_charge_mixed}
p_{q_g}\!\left(\boldsymbol\phi_{g'}^{(\mathcal L)}\right)
=
p_{q_g-\Delta q_g}\!\left(\boldsymbol\phi_{g'}\right),
\end{equation}
an intrinsic symmetry property \cite{FNentanglement} that follows directly from \eqref{eq:flux_charge} via
$\hat{\mathcal F}_{\mathcal L}^{-1}\hat P_{q_g}\hat{\mathcal F}_{\mathcal L}=\hat P_{q_g-\Delta q_g}$,
with
$\Delta q_g=\frac{\Delta\mathcal S_g}{2\pi/\mathcal N_g}$
the net change in $g$-charge.

So far the construction applies to general Abelian symmetries. In what follows we illustrate it with two classes of examples: (i) fermionic models with $U(1)$ and chiral $U_\chi(1)$ symmetries, and (ii) spin models with discrete symmetries.

\textit{\color{red}{Example 1: A fermion model with continuous symmetries.--}}
We illustrate symmetry-charge flow and its non-perturbative consequences in a fermionic model with continuous symmetries. Specifically, we consider staggered fermions \cite{kogut1975prd,banks1976prd,susskind1977prd}, a lattice regularization of a massless Dirac fermion, at inverse temperature $\beta$,
\begin{equation}\label{eq:H_sf}
\hat\rho=e^{-\beta \hat H_{\mathrm{sf}}},
\qquad
\hat H_{\mathrm{sf}}=\sum_{j=1}^{N}\hat h_{j+\frac{1}{2}},
\end{equation}
with $\qquad
\hat h_{j+\frac{1}{2}}=-i\!\left(\hat\psi_j^\dagger \hat\psi_{j+1}-\hat\psi_{j+1}^\dagger \hat\psi_j\right)$, and $\hat\psi_j$ ($\hat\psi_j^\dagger$) the fermion annihilating (creating) operator on site $j$.
In momentum space,
$\hat H_{\mathrm{sf}}=\sum_k 2\sin(k)\,\hat\psi_k^\dagger \hat\psi_k$ has zero modes at $k=0,\pi$. Expanding around these points yields right- and left-movers, reproducing Weyl fermions of opposite chirality in the low-energy limit.

The model has $U(1)\times U_\chi(1)$ symmetry generated by $\hat{\boldsymbol Q}\equiv(\hat Q,\hat Q_\chi)$ \cite{chatterjee2025prl},
\begin{equation}\label{eq:lattice_chiral}
\begin{cases}
\hat Q=\sum_i \hat\psi_i^\dagger \hat\psi_i,\\[2pt]
\hat Q_\chi=\frac{1}{2}\sum_{j=1}^{N}\!\left[\left(\hat\psi_j+\hat\psi_j^\dagger\right)\left(\hat\psi_{j+1}-\hat\psi_{j+1}^\dagger\right)+1\right],
\end{cases}
\end{equation}
with integer eigenvalues $(q,q_\chi)$.
Notably, $\hat Q_\chi$ is non-onsite, but becomes local in the low-energy limit and reproduces the chiral charge,
$\hat Q_\chi \to \frac{1}{2}\sum_{\delta k}\!\left(
\hat\psi_{\delta k}^\dagger \hat\psi_{\delta k}
-\hat\psi_{\pi+\delta k}^\dagger \hat\psi_{\pi+\delta k}
+1\right)$ for $\delta k\ll 1$ \cite{chatterjee2025prl}.

\begin{figure}[t]
\centering
\includegraphics[scale=0.18]{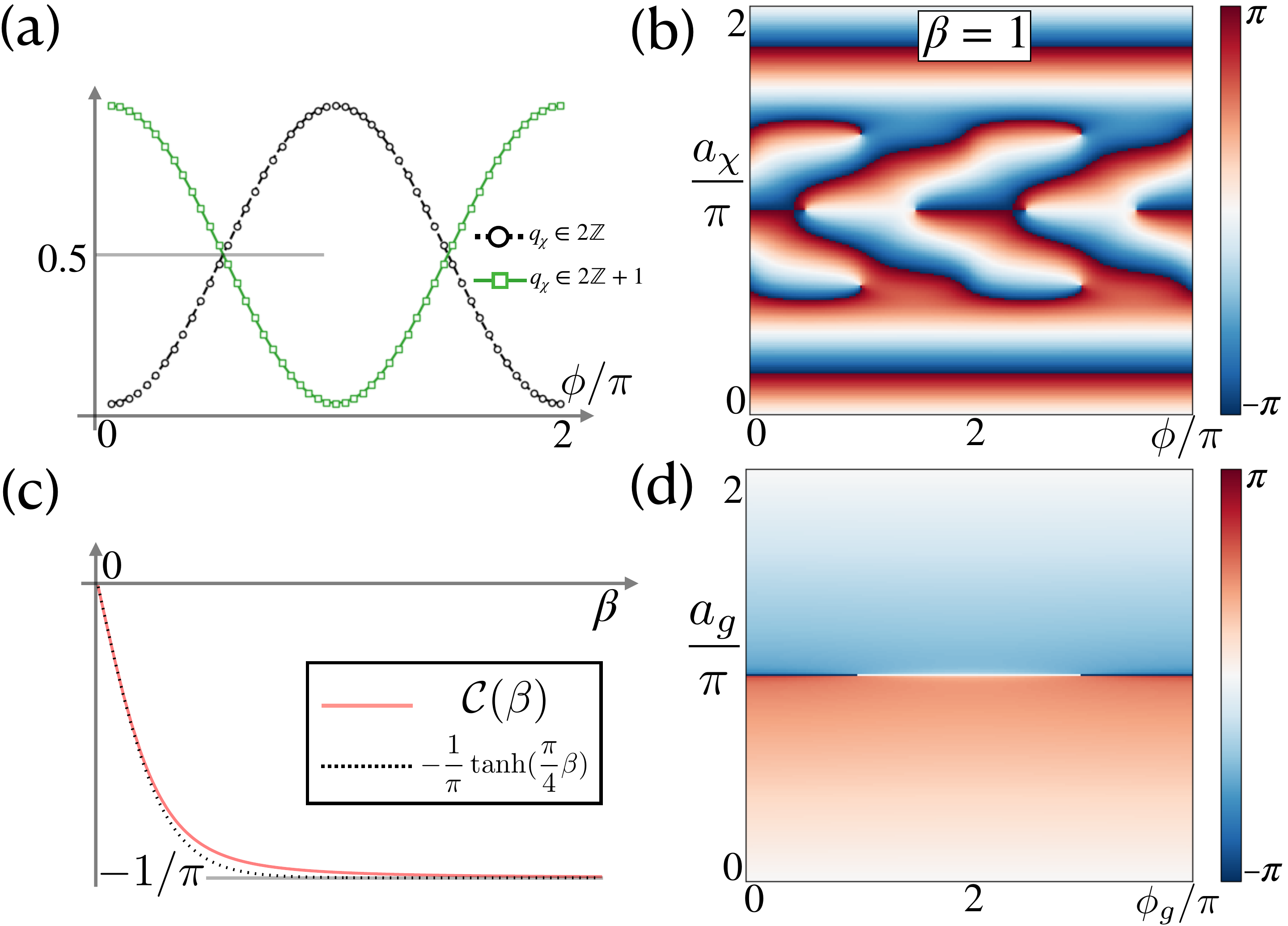}

\caption{Staggered fermions (a--c) and a $\mathbb{Z}_2$ spin model (d).
(a) Sector exchange under flux insertion: $p_{q_{\chi}\in2\mathbb{Z}}$ and $p_{q_{\chi}\in2\mathbb{Z}+1}$ versus $\phi=\oint A_x\,dx$, at $\beta=1$ for a $10$-site system.
(b) $\mathcal S_\chi$ versus $a_\chi$ and $\phi$.
(c) $\mathcal C(\beta)$ (red solid) versus $\beta$, compared with $-\frac{1}{\pi}\tanh\!\left(\frac{\beta\Lambda}{2}\right)$ \cite{huang2022prb} at $\Lambda=\frac{\pi}{2}$ (dotted); the $\Lambda\to\infty$ limit reproduces the result from dimensional regularization or the Fujikawa method \cite{das1997ws}.
(d) $\mathbb{Z}_2$ symmetry: $\arg\mathrm{Tr}\!\left[\hat\rho(\boldsymbol\phi_g)\,(\hat U_g)^{\boldsymbol a_g/\pi}\right]$ versus $\boldsymbol\phi_g\in[0,4\pi]$ and $\boldsymbol a_g\in[0,2\pi]$ (12 sites), showing quantized phase evolution under large gauge transformations $\boldsymbol\phi_g\rightarrow \boldsymbol\phi_g+2\pi$.
For visual clarity, $\boldsymbol\phi_g$ is treated as a continuous interpolation parameter between $0$ and $4\pi$.
Additionally, $\mathrm{Tr}(\hat\rho \hat U_g)$ vanishes for odd $N$ due to an extra anti-unitary symmetry $\hat{\mathcal V}=(\prod_{i\in\mathrm{even}}\sigma_i^x)\mathcal K$, with $\mathcal K$ complex conjugation \cite{FNau}.
\label{fig:masslessDirac_finiteT}}
\end{figure}

To connect with the anomaly equation \eqref{eq:chiral_anomaly}, we study the flow in the $(0,q_\chi)$ sector induced by a $U(1)$ spatial flux $\boldsymbol\phi_g=(\phi,0)$ (henceforth writing $\phi$ and $q_\chi$ when unambiguous).
The corresponding flux-insertion operator $\hat{\mathcal F}(\phi)$ satisfies (see \cite{FN_sm}),
\begin{equation}\label{eq:flux_chi_sym}
\left[\hat{\mathcal F}(n\pi)\right]^{-1} e^{i\pi \hat Q_\chi}\,\hat{\mathcal F}(n\pi)
=
(-1)^n\, e^{i\pi \hat Q_\chi},
\qquad
n\in\mathbb Z .
\end{equation}
The extra factor $(-1)^n$ is a direct signature of the non-onsite character of $\hat Q_\chi$, as it would be absent for an onsite symmetry.
Equation~\eqref{eq:flux_chi_sym} implies that a $\pi$ flux exchanges even/odd chiral-charge sectors [Fig.~\ref{fig:masslessDirac_finiteT}(a)],
\begin{equation}\label{eq:chi_sector_swap}
p_{q_\chi \!\!\!\!\!\mod 2}(\phi+n\pi)
=
p_{q_\chi \!\!\!\!\!\mod 2-\Delta q_\chi}(\phi),
\qquad
\Delta q_\chi = n \ \mathrm{mod}\ 2 .
\end{equation}
This $\mathbb{Z}_2$ structure follows from the non-commutativity of $\hat{Q}$ and $\hat{Q}_\chi$, but $[\hat{Q}, (-1)^{\hat{Q}_\chi}]=0$, leaving a $\mathbb{Z}_2$ chiral symmetry.
Crucially, the resulting anomaly is insensitive to state purity. Taking the large gauge transformation to be $\hat{\mathcal F}_{\mathcal L}=\hat{\mathcal F}(2\pi)$, Eq.~\eqref{eq:flux_chi_sym} yields $\Delta\mathcal S_g\in 2\pi\mathbb Z$ for the symmetry $g=(-1)^{\hat{Q}_\chi}$, i.e., an even shift of chiral charge. In the zero-temperature limit this reduces to the familiar spectral-flow statement that inserting one flux quantum shifts the ground-state chiral charge by $-2$.
The symmetry-charge flow further imposes non-perturbative constraints on the effective action. Writing $\mathcal S_\chi$ as the phase of the flux-inserted partition function, $\mathcal S_\chi\equiv \arg Z[a_\chi,\phi]$, Eq.~\eqref{eq:flux_chi_sym} requires that at $a_\chi=\pi$ the phase changes by an odd multiple of $\pi$ under insertion of a half flux quantum,
$\frac{1}{\pi}\,\mathcal S_\chi\big|_{\phi}^{\phi+\pi}\in 2\mathbb Z+1.$ This constitutes a mixed-state generalization of anomaly inflow~\cite{callan1985npb}, as it reduces to the Chern-Simons term in the zero-temperature limit:
Keeping only terms linear in the gauge fields \cite{deser1997prl,dunne1997prl} gives
$\mathcal S_\chi=\frac{\mathrm{ch}}{\pi}a_\chi\phi$, with $\mathrm{ch}\in 2\mathbb Z+1$ fixed by the flow above, where $a_\chi=\oint A^\chi$ is the temporal $U_\chi(1)$ flux;
viewing the spatial manifold $S^1$ as the boundary of a disk $D$ makes the corresponding Chern--Simons form explicit,
$\mathcal S_\chi=\frac{\mathrm{ch}}{\pi}\int_{S^1\times D}\epsilon^{\mu\nu\rho}A^\chi_\mu \partial_\nu A_\rho$,
so that the boundary anomaly on $S^1$ is canceled by bulk inflow \cite{callan1985npb}.
We confirm these predictions numerically in Fig.~\ref{fig:masslessDirac_finiteT}(b).

For comparison, we compute the chiral current non-conservation Eq.~\eqref{eq:chiral_anomaly}, which generically receives thermal contributions. Using the lattice chiral-symmetry operator in Eq.~\eqref{eq:lattice_chiral},
we obtain, to linear order in $A$ and in the thermodynamic limit~\cite{FN_sm},
\begin{equation}\label{eq:chiralA_finiteT}
\partial_\mu \langle \hat j_\chi^\mu\rangle
=
\mathcal C(\beta)\,\epsilon^{\mu\nu}\partial_\mu A_\nu .
\end{equation}
The coefficient $\mathcal C(\beta)$ is non-universal and depends on microscopic details [Fig.~\ref{fig:masslessDirac_finiteT}(c)]:
for $\tfrac{\pi}{4}\beta\gg 1$, $\mathcal C(\beta)=-\tfrac{1}{\pi}$, matching the continuum result \eqref{eq:chiral_anomaly}, whereas for $\tfrac{\pi}{4}\beta\ll 1$ one finds $\mathcal C(\beta)\sim -\tfrac{1}{4}\beta$.
Thus, unlike symmetry-charge flow, current non-conservation need not furnish a robust finite-temperature anomaly diagnostic.

\textit{\color{red}{Example 2: Bosonic models with discrete symmetries.--}}
We now turn to bosonic models with discrete symmetries. We first illustrate the construction for the $\mathbb Z_2$-symmetric spin chain of Refs.~\cite{levin2012prb,xu2025arxiv}, and then comment on the general Abelian case. It is convenient to write 
$\hat\rho=e^{-\hat{K}}$. The corresponding flux-inserted $\hat{K}$ is then,
\begin{eqnarray}\label{eq:Ham_Z2}
\hat K(\boldsymbol\phi_g)
&=&
-\sum_{i=2}^{2N}\Big(\sigma_i^x-\sigma_{i-1}^z\sigma_i^x\sigma_{i+1}^z\Big)\nonumber\\
&&
-e^{-i\frac{1}{2}\boldsymbol\phi_g\,\sigma_1^z\sigma_{2N}^z}\Big(\sigma_1^x-\sigma_{2N}^z\sigma_1^x\sigma_2^z\Big),
\end{eqnarray}
where $\sigma_i^a$ are Pauli matrices and, for simplicity, we take a chain of length $2N$.
The $\mathbb Z_2$ symmetry is generated by
$\hat U_g=\left(\prod_i \sigma_i^x\right)\prod_i\!\left[e^{i\frac{\pi}{4}(1+\sigma_i^z\sigma_{i+1}^z)}\right]$.
Following Eq.~\eqref{eq:flux_insertion}, the flux-insertion operator $\hat{\mathcal F}(\boldsymbol\phi_g)$ is constructed from $\hat U_g$, and \eqref{eq:Ham_Z2} then follows by conjugation.
The allowed flux values are $\boldsymbol\phi_g\in\{0,\pi\}$, as required by $\mathbb Z_2$.
A large gauge transformation corresponds to $\boldsymbol\phi_g^{(\mathcal L)}=\boldsymbol\phi_g+2\pi$ and is generated by the local operator $\hat{\mathcal F}_{\mathcal L}=\sigma_1^z$.
One then finds
$\hat{\mathcal F}_{\mathcal L}^{-1}\hat U_g \hat{\mathcal F}_{\mathcal L}=-\hat U_g$,
realizing symmetry-charge flow via Eq.~\eqref{eq:flux_charge}.
We confirm this numerically in Fig.~\ref{fig:masslessDirac_finiteT}(d).

More generally, symmetry-charge flow occurs in one-dimensional bosonic systems with Abelian symmetries, including reduced density matrices \cite{cirac2011prb} of two-dimensional SPT phases \cite{chen2011science,schuch2011prb}.
In these models, the symmetry operators and the large gauge transformation obey Eq.~\eqref{eq:flux_charge} (see Appendix~\ref{supp_sec:flux_tn} for a tensor-network derivation), with $e^{i\Delta\mathcal S_g}$ given by a group 3-cocycle and $\hat{\mathcal F}_{\mathcal L}$ fixed by the flux-composition rule
$\hat{\mathcal F}(\boldsymbol\phi_{g_1})\,\hat{\mathcal F}(\boldsymbol\phi_{g_2})
=
\hat{\mathcal F}_{\mathcal L}\,\hat{\mathcal F}(\boldsymbol\phi_{g_1g_2}).$

\textit{\color{red}{Conclusions and outlook.}--}
Our results demonstrate 't~Hooft anomalies as universal, non-perturbative diagnostics for mixed states. Extending existing classifications, our formulation applies whether or not symmetry charge can be exchanged with the environment, and also captures mixed-state anomalies arising from reduced density matrices of SPT phases. While we focus here on anomaly constraints imposed on states, a natural next step is to analyze constraints on dynamics. Looking ahead, these may sharpen symmetry-based characterizations of mixed-state phases and enable anomaly-guided state preparation.

\begin{acknowledgments}
We thank Bo Han, Frank Pollmann, and Xiao-Qi Sun for discussions. Z.-M.~H. and S.~D. are supported by the  Deutsche Forschungsgemeinschaft (DFG, German Research Foundation) under Germany's Excellence Strategy Cluster of Excellence Matter and Light for Quantum Computing (ML4Q) EXC 2004/1 390534769 and by the DFG Collaborative Research Center (CRC) 183 Project No. 277101999 - project B02.
\end{acknowledgments}

\begin{appendix}

\clearpage
\section*{End Matter}

\section{Flux insertion and large gauge transformations on finite periodic chains \label{supp_sec:flux_insertion}}
\begin{figure}[htbp]
\includegraphics[scale=0.018]{appfig_flux.pdf}
\caption{Illustration of flux insertion on a finite periodic chain. \label{appfig:flux}}
\end{figure}

Here we detail the flux-insertion operator on an $N$-site periodic chain for a general Abelian symmetry group $G$, complementing the infinite-chain result in the main text [Eq.~\eqref{eq:flux_insertion}]. For a local symmetry operator $\hat{U}_g$, i.e., realizable by a unitary circuit of depth much smaller than $N$, we demonstrate that the large-gauge transformation operator $\hat{\mathcal{F}}_{\mathcal{L}}$ remains a local unitary on the finite ring. Consequently, results in the main text, which rely only on $\hat{\mathcal{F}}_{\mathcal{L}}$, remain valid at finite size. However, a caveat is that a generic flux insertion for a finite periodic chain need not be realizable as a similarity transformation, since the untwisted and flux-threaded density matrices need not be isospectral, as discussed below.

To this end, we first review the usual implementation of flux insertion for a periodic $N$-site chain. Let $\hat{G}^{\left(N\right)}\equiv-\ln\hat{\rho}^{\left(N\right)}$
be a local modular Hamiltonian, where the superscript $N$ emphasizes the finite system size. This construction uses locality to implement the flux directly at the level of the modular Hamiltonian, while the more general construction presented below does not require this it.
Specifically, choosing a cut, we decompose [see Fig. \ref{appfig:flux}(a)]
\begin{equation}
\hat{G}^{\left(N\right)}=\hat{G}^{\left(N\right)}_{L}+\hat{G}^{\left(N\right)}_{R}+\hat{G}^{\left(N\right)}_{LR}+\hat{G}^{\left(N\right)}_{RL}.
\end{equation}
The first two terms are supported within the left and right intervals, while the last two cross the corresponding boundaries. A $g$-flux insertion gives
\begin{equation}
\hat{G}^{\left(N\right)}\left(\boldsymbol{\phi}_{g}\right)=\hat{G}^{\left(N\right)}_{L}+\hat{G}^{\left(N\right)}_{R}+\hat{U}_{g,\ R}\hat{G}^{\left(N\right)}_{LR}\hat{U}^{\dagger}_{g,\ R}+\hat{G}^{\left(N\right)}_{RL},\label{app_eq:modular_Ham_finite}
\end{equation}
where $\hat{U}_{g,R}$ acts only on the right-supported degrees of freedom of $\hat{G}^{\left(N\right)}_{LR}$. In general, the resulting $\hat{\rho}^{\left(N\right)}\left(\boldsymbol{\phi}_{g}\right)$
and $\hat{\rho}^{\left(N\right)}\left(0\right)$ need not be isospectral.

Alternatively, we implement this construction by viewing the finite periodic chain as a length-$N$ unit cell of a translationally invariant infinite chain and identifying its two ends, i.e., taking the quotient $\mathbb{Z}/(N\mathbb{Z})$~\cite{zaletel2013prl,zaletel2014njp,huang2025prl_mspt} [see Fig.~\ref{appfig:flux}(b)].-- This construction has two advantages: (i) it applies to a general density matrix, irrespective of its microscopic form; (ii) it makes the large-gauge-transformation operator explicit.
Concretely, let
$\hat U_g\equiv e^{i\boldsymbol\phi_g\cdot\hat{\boldsymbol Q}}$ be the symmetry operator of the parent infinite chain. A g-flux at $x_0$ is inserted by acting with the unitary restriction of the symmetry operator to the half-infinite interval $[x_0, \infty)$,
$\hat{U}_{g, [x_0, \infty)}\equiv e^{i\boldsymbol\phi_g\cdot\hat{\boldsymbol Q}_{[x_0,\infty)}}$, whose unitarity follows from the finite-depth nature of $\hat{U}_g$.
To obtain an $N$-periodic flux pattern, we repeat this operation every $N$ sites, $x_0\to x_0+mN$
[Fig.~\ref{appfig:flux}(c)]. Identifying the ends of the unit cell therefore yields an $N$-site periodic chain with a $g$-flux inserted.

As a consistency check, we now specialize to the case in which the parent infinite-chain density matrix admits a local modular Hamiltonian, $\hat{G}^{\left(\infty\right)}=\sum_{i}\hat{G}^{\left(\infty\right)}_{i}$. This locality assumption is used only to verify that the quotient construction above reproduces the conventional finite-ring prescription in Eq.~\eqref{app_eq:modular_Ham_finite}, while it is not required to define the periodic flux insertion or the associated large-gauge-transformation operator. We then choose a cut at $x_{0}$, and the unitary restriction
$\hat{U}_{g,[x_0,\infty)}$ of symmetry $\hat{U}_{g}$ modifies only terms supported within a finite neighborhood of the cut; all sufficiently distant terms remain unchanged. Applying the periodic flux-insertion operator $\hat{\mathcal{F}}^{\left(\infty\right)}$ shown in Fig.~\ref{appfig:flux}(c) therefore gives,

\begin{eqnarray}
\hat{G}^{(\infty)}(\boldsymbol{\phi}_g) & \equiv & \hat{\mathcal{F}}^{\left(\infty\right)}\hat{G}^{\left(\infty\right)}\left(0\right)\hat{\mathcal{F}}^{\left(\infty\right)-1}\nonumber \\
 & = & \sum_{m\in\mathbb{Z}}\sum_{i:\,\mathrm{supp}(\hat{G}_{i})\cap\mathcal{B}_{m}=\varnothing}\hat{G}^{(\infty)}_{i}\nonumber \\
 && +\sum_{m\in\mathbb{Z}}\sum_{\substack{i:\,\mathrm{supp}(\hat{G}_{i})\cap\mathcal{B}_{m}\neq\varnothing}} \mathrm{Ad}_{g, m}[\hat{G}^{\infty}_i],
\end{eqnarray}
where $\mathrm{Ad}_{g, m}[\hat{G}^{\infty}_i]\equiv \hat{U}_{g,[x_{0}+mN,\infty)}\left[\hat{G}^{(\infty)}_{i}\right]\hat{U}^{\dagger}_{g,[x_{0}+mN,\infty)}$, and
$\mathcal{B}_{m}$ denotes a finite neighborhood of the cut
$x_{0}+mN$. This infinite chain contains one $g$-twist per unit cell, so quotienting by translation of $N$ lattice sites [$\mathbb{Z}/N\mathbb{Z}$, see Fig.~\ref{appfig:flux}(b)] yields precisely the finite-ring result $\hat{G}^{(N)}(g)$ defined above. 

However, following this construction, for a generic flux insertion, the infinite-chain operator $\hat{\mathcal F}^{(\infty)}(\boldsymbol{\phi}_g)$ need not descend to an operator on a single $N$-site unit cell, making its finite-size quotient generally nontrivial. Crucially, the large-gauge transformation $\hat{\mathcal F}_{\mathcal L}^{(\infty)}$, on which our results rest, does descend. Indeed, $\hat{\mathcal{F}}^{\left(\infty\right)}_{\mathcal{L}}\equiv\hat{\mathcal{F}}^{\left(\infty\right)}\left(\boldsymbol{\phi}_{g_{1}}\right)\hat{\mathcal{F}}^{\left(\infty\right)}\left(\boldsymbol{\phi}_{g_{2}}\right)\left[\hat{\mathcal{F}}^{\left(\infty\right)}\left(\boldsymbol{\phi}_{g_{1}g_{2}}\right)\right]^{-1}$ is localized near the cut and therefore renders a well-defined local unitary on the finite ring. This localization follows from the finite-depth locality of the symmetry operator $\hat U_g$: the factors in the composition cancel everywhere sufficiently far from the cut. Consequently, once $N$ exceeds the circuit depth of $\hat{U}_g$, its translated copies do not overlap, and it descends to a local unitary $\hat{\mathcal F}_{\mathcal L}^{(N)}$ on the $N$-site ring.

\section{Large-gauge transformation operator in tensor networks \label{supp_sec:flux_tn}}
We show that mixed-state anomalies and symmetry-charge flow arise in general bosonic models. To this end, using tensor-network methods on a finite periodic chain, we construct the operator implementing large gauge transformations and derive Eq.~\eqref{eq:flux_charge} for these systems.

We focus on symmetries arising from reduced density matrices of two-dimensional SPT phases; see Ref.~\cite{FN_sm} for a detailed derivation based on projected entangled-pair state (PEPS) representations of SPTs~\cite{willamson2016prb, jiang2017prb, garrerubio2023quantum}.
Specifically, for a periodic chain, the symmetry operator admits a matrix-product-operator (MPO) representation \cite{FN_sm},
\begin{equation}
\hat{U}_{g}=\begin{array}{c}
	\begin{tikzpicture}[scale=.4, baseline={([yshift=0ex]current bounding box.center)}, thick]
		\ATensorMPO{0, 1}{$A_g$}
        \ATensorMPO{2, 1}{$A_g$}
        \draw[shift={(0,0)},dotted] (2.8,1) -- (5,1);
        \ATensorMPO{6, 1}{$A_g$}
        \draw[shift={(0,2.8)}] (-1.2,-1.8) arc(90:270:0.2);
         \draw[shift={(0,0)}] (7.2,0.6) arc(-90:90:0.2);
	\end{tikzpicture}\end{array}, 
\end{equation}
where \begin{equation}
 \begin{array}{c}
  \begin{tikzpicture}[scale=0.4, baseline={([yshift=0ex]current bounding box.center)}, thick]
      \ATensorMPO{2, 1}{$A_g$}
  \end{tikzpicture}
  \end{array}=
  \begin{array}{c}
  \begin{tikzpicture}[scale=0.4, baseline={([yshift=0ex]current bounding box.center)}, thick]
       \draw (0, -1)--(0, 1);
       \draw (-0.75, 0)--(0, 0);
       \draw (0.75, 0)--(0, 0);
       \wTensor{0, 0.5}{$R_g$}
       \xTensor{0.9, 0}{\scalebox{1}{$\alpha_g$}}

       \draw (0.5,-1.2)   node {\scriptsize $g_1$};
       \draw (0.8,2)   node {\scriptsize $g_1^\prime$};

       \draw (-1,-0.5)   node {\scriptsize $g_l$};
       \draw (1.8,-0.5)   node {\scriptsize $g_r$};
  \end{tikzpicture}
  \end{array}= \delta_{g_1^\prime, g_1g^{\scalebox{0.5}{$-1$}}}\delta_{g_1, g_l} \alpha(g_1 g_r^{-1}, g_r g^{-1}, g).
\end{equation}
The physical degree of freedom carries the regular representation of $G$, and is therefore labeled by group elements. $\alpha_g$ acts diagonally,
$\alpha_g |g_l, g_r\rangle\equiv \alpha[g_lg_r^{-1}, R_g (g_r), g]\ |g_l, g_r\rangle$,
where $R_g$ denotes the right action, $R_g|g_1\rangle=|g_1g^{-1}\rangle$.
Here $\alpha: G\times G\times G\to U(1)$ is a $3$-cocycle in standard form (see \cite{FN_sm} for details), and satisfies the pentagon equation, 
 \begin{equation}
\frac{\alpha(g_{12},g_3, g_4)\alpha(g_1,g_2,g_{34})}{\alpha(g_1,g_2, g_3)\alpha(g_1,g_{23}, g_4)\alpha(g_2,g_3, g_4)}=1.
\end{equation}
Meanwhile, other forms are related by a shallow-depth unitary (see \cite{FN_sm} for a brief review).

Below, we derive the operator implementing the large gauge transformation associated with $\hat U_{g_1}\hat U_{g_2}\hat U_{g_1g_2}^\dagger$ [see Eq.~\eqref{supp_eq:flux_abelian} for the result], denoted $\hat{\mathcal F}_{\mathcal L_{1,2}}$ for clarity.

\paragraph{Flux-insertion operator.--} The flux-insertion operator follows from the composition rules of symmetry operators with open boundary conditions. We therefore first write the open-boundary form of $\hat U_g$,
\begin{equation}
\hat{U}_g^{(\text{o})}=
\begin{array}{c}
  \begin{tikzpicture}[scale=0.4, baseline={([yshift=-1.9ex]current bounding box.center)}, thick]
       \draw (-1, -1)--(-1, 1);
       \draw (1, -1)--(1, 1);
       \draw (8, 0)--(8.5, 0);
       \draw[dotted](8.5, 0)--(9.5, 0);
       
       \xTensor{0, 0}{$\alpha_g$}
       \wTensor{-1, 0.5}{$R_g$}
       \wTensor{1, 0.5}{$R_g$}

        \xTensor{2, 0}{$\alpha_g$}

         \draw (3, -1)--(3, 1);
         \wTensor{3, 0.5}{$R_g$}

         \draw[dotted](3, 0)--(6, 0);
         \draw(3, 0)--(3.5, 0);
         \draw(5.5, 0)--(6, 0);

         \draw (6, -1)--(6, 1);
         \wTensor{6, 0.5}{$R_g$}
          \xTensor{7, 0}{$\alpha_g$}
           \draw (8, -1)--(8, 1);
         \wTensor{8, 0.5}{$R_g$}
  \end{tikzpicture}
  \end{array},
\end{equation}
with the boundary link removed. 

For later convenience, we introduce the following operator,
\begin{equation}\label{supp_eq:w_g}
w_g=
 \begin{array}{c}
  \begin{tikzpicture}[scale=0.4, baseline={([yshift=-1.9ex]current bounding box.center)}, thick]
       \draw (-1, -1)--(-1, 1);
       \draw (1, -1)--(1, 1);
       \xTensor{0, 0}{$\alpha_g$}
       \wTensor{-1, 0.5}{$R_g$}
       \wTensor{1, 0.5}{$R_g$}
  \end{tikzpicture}
  \end{array},
\end{equation}
which serves as the building block of $\hat{U}_g^{(\text{o})}$.
We also define $d_{1,2}\equiv w_{g_1}w_{g_2}w_{g_1g_2}^\dagger$, which satisfies
\begin{eqnarray}\label{supp_eq:d_mul}
&&d_{1,2}|g_l, g_r\rangle\nonumber\\
{}&=& \alpha(g_l, g_1, g_2)\times \alpha(g_r, g_1, g_2)^* |g_l, g_r\rangle.
\end{eqnarray}
This follows from the pentagon equation applied to $(g_l g_r^{-1},\,g_r,\,g_1,\,g_2)$.
Notably, $d_{1,2}$ is diagonal in the two-site product basis $|g_l\rangle\otimes|g_r\rangle$.

Following Appendix~\ref{supp_sec:flux_insertion}, the flux-insertion operator is inferred from the multiplication of open-boundary symmetry operators. In particular, using Eq.~\eqref{supp_eq:d_mul} and that $\alpha_g$ is diagonal in the $|g_l,g_r\rangle$ basis, we find
\begin{equation}
\hat U_{g_1}^{(\mathrm{o})}\hat U_{g_2}^{(\mathrm{o})}\hat U_{g_1g_2}^{(\mathrm{o})^\dagger}
=
\hat{\mathcal F}_{\mathcal L_{1,2}}\big|_{i=1}\times(\cdots)\big|_{i\to\infty}.
\end{equation}
Here $i=1$ labels the beginning of the cut [see Fig.~\ref{appfig:flux}(c) in Appendix~\ref{supp_sec:flux_insertion}], while the contribution at $i\to\infty$ drops out. Repeating the construction under translations by $N$ sites gives one copy of $\hat{\mathcal F}_{\mathcal L_{1,2}}$ per unit cell, so quotienting by these translations yields the corresponding operator on the finite periodic chain, which obeys
\begin{equation}\label{supp_eq:flux_abelian}
\hat{\mathcal{F}}_{\mathcal{L}_{1,2}}|g\rangle= \alpha(g, g_1, g_2)|g\rangle.
\end{equation}
Crucially, it satisfies the following algebraic relation,
\begin{equation}
\hat{\mathcal{F}}_{\mathcal{L}_{1,2}}^{-1} \hat{U}_g \hat{\mathcal{F}}_{\mathcal{L}_{1,2}}|g_0, \dots \rangle=  \frac{\alpha(g_0, g_1, g_2)}{\alpha(g_0 g^{-1}, g_1, g_2)} \hat{U}_g|g_0, \dots\rangle,
\end{equation}
which for an Abelian symmetry with $\alpha$ in the standard form, reduces to
\begin{equation}
\frac{\alpha(g_0, g_1, g_2)}{\alpha(g_0g^{-1}, g_1, g_2)} =1/\alpha(g^{-1}, g_1, g_2) \in U(1),
\end{equation}
reproducing Eq.~\eqref{eq:flux_charge} in the main text.

\end{appendix}

\begin{thebibliography}{66}
\bibitem [{FNtrace()}]{FNtrace}%
  \BibitemOpen
  \href@noop {} {}\bibinfo {note} {Here, trace identities in the thermodynamic
  limit are defined through finite-size regularization: we first work on a
  finite $N$-site lattice, where the trace is well defined, and only then take
  the limit $N\to\infty$.}\BibitemShut {Stop}%
\bibitem [{FNm()}]{FNmath}%
  \BibitemOpen
  \href@noop {} {}\bibinfo {note} {This captures the anomaly within the
  framework of~\cite{tachikawa2020sp}, where the anomaly can be characterized
  by a finite Abelian group extension.}\BibitemShut {Stop}%
\bibitem [{FNa()}]{FNau}%
  \BibitemOpen
  \href@noop {} {}\bibinfo {note} {This follows from anti-unitary symmetry and
  Hermiticity ($\hat{U}_g=\hat{U}_g^\dagger$), which imply
  $\text{Tr}\left(\hat{U}_{g}\hat{\rho}\right)=(-)^{N}\text{Tr}\left(\hat{U}_{g}^\dagger\hat{\rho}\right)$,
  and thus $\text{Tr}\left(\hat{U}_{g}\hat{\rho}\right)=0$ when $N$ is
  odd.}\BibitemShut {Stop}%
\end{thebibliography}
\end{document}


\title{Supplemental Material for "Symmetry-charge flow from mixed-state anomaly"}
\author{Ze-Min Huang }


\author{Sebastian Diehl}

\affiliation{Institute for Theoretical Physics, University of Cologne, 50937 Cologne, Germany}

\begin{abstract}
This supplemental material includes details on: (i) How symmetry-charge flow implies long-range entanglement; (ii) An illustration of symmetry-charge flow for Abelian symmetries using the $K$-matrix framework; (iii) A derivation of the anomaly equation in Eq.~\eqref{eq:chiralA_finiteT}, and the relation between the zero-temperature current algebra and symmetry-charge flow; (iv) A derivation of the algebraic relation between the flux-insertion and symmetry operators for staggered fermions, $\left[\hat{\mathcal{F}}\left(\pi\right)\right]^{-1}e^{i\pi\hat{Q}_{\chi}}\hat{\mathcal{F}}\left(\pi\right)=\left(-1\right)e^{i\pi\hat{Q}_{\chi}}$; (v) A derivation of the reduced density matrices and symmetry operators in matrix-product form for a two-dimensional SPT state; (vi) A summary of useful properties of $3$ cocycle.

\end{abstract}

\maketitle

\clearpage 

\setcounter{equation}{0}
\setcounter{figure}{0}
\setcounter{table}{0}
\setcounter{page}{1}
\setcounter{section}{0}
\makeatletter
\renewcommand{\theequation}{S\arabic{equation}}
\renewcommand{\thefigure}{S\arabic{figure}}
\renewcommand{\thesection}{S\arabic{section}}
\renewcommand{\bibnumfmt}[1]{[S#1]}
\renewcommand{\citenumfont}[1]{S#1}

\tableofcontents

\section{Long-range entanglement in $\hat{\rho}_{q_g}$ \label{supp_sec:LRE_canonical}}

We demonstrate that $\hat{\rho}_{q_g}$, which has a fixed $g$-charge for each $g\in G$ and exhibits symmetry-charge flow, contains long-range entanglement, by showing that no purification can be short-range entangled.
Specifically, for mixed states, long-range entanglement means the state cannot be mapped to a product state by any finite-depth local quantum channel~\cite{hastings2011prl}; otherwise it is short-range entangled. Here, a local quantum channel is one induced by a finite-depth local circuit acting on a Hilbert space $\mathcal{H}_S\otimes \mathcal{H}_A$, where $\mathcal{H}_A$ is a finite-dimensional ancilla attached to each onsite physical Hilbert space of $\mathcal{H}_S$.
When applied to a density matrix with a local modular Hamiltonian, symmetry-charge flow requires the modular Hamiltonian to have a degenerate ground-state subspace.-- By contrast, if the ground state were unique, it would simultaneously have a fixed $g$-charge and be short-range entangled, a contradiction. 

Below, we shall first consider bosonic models with discrete symmetries, then extend this to fermionic models with chiral symmetry-charge flow.

\subsection{Bosonic models with discrete symmetries}
Concretely, we consider the following symmetry actions on the fixed $g$-charge sector,
\begin{equation}\label{supp_eq:sym_densityM}
\hat{U}_{g}\hat{\rho}_{q_g}=e^{i \frac{2\pi}{\mathcal{N}_g} q_g}\hat{\rho}_{q_g}\quad  \forall g \in G,
\end{equation}
where $\hat{U}_g \equiv e^{i \boldsymbol{a}_g \cdot \hat{\boldsymbol{Q}}}$.
Additionally, we assume that $\hat{U}_g$ admits an onsite realization on an enlarged Hilbert space, i.e., 
\begin{equation}\label{supp_eq:sym_enlarged}
\hat{U}_g=w_g^{\otimes N} \hat{P}_e,
\end{equation}
where $w_g$ is a local operator, and $\hat{P}_e$ projects onto the original (physical) Hilbert space. This assumption holds for all symmetries considered in the main text (see Sec.~\ref{supp_sec:onsite_rep} for a demonstration).  We expect it to extend to anomalies when $\hat{\mathcal{F}}_L$ is Abelian, although a general proof lies beyond our scope. This is because any finite-group 't Hooft anomaly admits a finite Abelian extension $0\rightarrow A \rightarrow \Gamma \rightarrow G\rightarrow 0$, with $A$ generated by $\hat{\mathcal{F}}_L$, such that $\Gamma$ is non-anomalous \cite{wang2018prx,tachikawa2020sp}. This suggests that $\Gamma$ admits an onsite representation in an enlarged Hilbert space.

A purification of $\hat\rho_{q_g}$ is a pure state $|\rho_{q_g}\rangle\in \mathcal{H}_S\otimes \mathcal{H}_A$  such that tracing out the ancilla reproduces $\hat\rho_{q_g}$, i.e.,
\begin{equation}
\hat\rho_{q_g}\equiv \text{Tr}_a |\rho_{q_g}\rangle  \langle \rho_{q_g}|, 
\end{equation}
where $\text{Tr}_a$ traces out the ancilla. Moreover, every purification has the following symmetry properties, 
\begin{equation}\label{supp_eq:symm_purification}
\hat{U}_{g}\otimes\mathbb{I}^{\left(a\right)}|\rho_{q_g}\rangle=e^{i\frac{2\pi}{\mathcal{N}_g}q_g}|\rho_{q_g}\rangle
\end{equation}
with the superscript $\left(a\right)$ for ancilla, which follows from Eq.~\eqref{supp_eq:sym_densityM}. Also, the onsite realization of the symmetry action on the enlarged Hilbert space [Eq.~\eqref{supp_eq:sym_enlarged}] implies that
\begin{equation}\label{supp_eq:onsite_sym}
\hat{U}_g\otimes \mathbb{I}^{\left(a\right)}|\rho_{q_g}\rangle =w_g^{\otimes N}\otimes \mathbb{I}^{\left(a\right)}|\rho_{q_g}\rangle,
\end{equation}
since,  by definition, $|\rho_{q_g}\rangle$, is invariant under the projection $\hat{P}_e$.

Assuming translation invariance, we then demonstrate that $|\rho_{q_g}\rangle$ cannot be short-range entangled. The key point is that for a short-range entangled pure state, the flux-insertion operator corresponding to  $\hat{U}_{g_1}\hat{U}_{g_2}\hat{U}_{g_1g_2}^\dagger$  must be trivial (i.e. proportional to the identity), and thus cannot generate symmetry-charge flow. 
The proof then proceeds by contradiction. Assuming that $|\rho_{q_g}\rangle$ is a short-range entangled state, then it admits a matrix-product-state representation~\cite{cirac2021rmp}.
\begin{equation}
|\rho_{q_g}\rangle=
	\begin{tikzpicture}[scale=.4, baseline={([yshift=-1.4ex]current bounding box.center)}, thick]
        \draw[shift={(0,0)},dotted] (2,0) -- (5,0);
    
		\ATensorMPS{0, 0}{$A$}
        \ATensorMPS{1.8, 0}{$A$}
        
        \ATensorMPS{5.7, 0}{$A$}
        
        \draw[shift={(0,0.8)}] (-1.2,-0.8) arc(90:270:0.2);
         \draw[shift={(-1,0)}] (7.8,-0.4) arc(-90:90:0.2);

	\end{tikzpicture}.
\end{equation}
Also, due to the fundamental theorem~\cite{cirac2021rmp}, the symmetries must be realized as similarity transformations in the virtual space, which can only form a projective representation. Namely, consider a state $|\rho^\prime_{q_g}\rangle$ 
\begin{equation}
|\rho^\prime_{q_g}\rangle=
	\begin{tikzpicture}[scale=.4, baseline={([yshift=-1.4ex]current bounding box.center)}, thick]
        \draw[shift={(0,0)},dotted] (2,0) -- (5,0);
    
		\ATensorMPS{0, 0}{$B$}
        \ATensorMPS{1.8, 0}{$B$}
        
        \ATensorMPS{5.7, 0}{$B$}
        
        \draw[shift={(0,0.8)}] (-1.2,-0.8) arc(90:270:0.2);
         \draw[shift={(-1,0)}] (7.8,-0.4) arc(-90:90:0.2);

	\end{tikzpicture}\ ,
\end{equation}
such that 
\begin{equation}
|\rho_{q_g}\rangle =e^{i\mathcal{N}\theta} |\rho^\prime_{q_g}\rangle,
\end{equation}
where $\mathcal{N}$ is the number of external legs introduced for later convenience. Also, without loss of generality, we take $B$ to be a normal tensor \cite{cirac2021rmp}, i.e., there exists a finite blocking length $L_0$ such that the tensor obtained by blocking $L_0$ consecutive sites defines an injective linear map from the virtual space to the physical space.
Then,
\begin{equation}
\begin{array}{c}
\begin{tikzpicture}[scale=.4, baseline={([yshift=-1.3ex]current bounding box.center)}, thick]
    \ATensorMPS{0,0}{$A$}
  \end{tikzpicture} \end{array}
  =e^{i\theta} 
  \begin{array}{c}
\begin{tikzpicture}[scale=.4, baseline={([yshift=-2.6ex]current bounding box.center)}, thick]
    \ATensorMPS{0,0}{$B$}
     \xTensor{1.6, 0}{$V$}
     \xTensor{-1.6, 0}{$V^{\scalebox{0.5}{-1}}$}
  \end{tikzpicture},
  \end{array}
  \label{supp_eq:MPS_injective_invertible}
\end{equation}
with $V$ an invertible matrix, implying that $|\psi_{q_g}\rangle$ and $|\psi_{q_g}^\prime\rangle$ can only differ by a boundary term. Applying to the symmetries in Eq.~\eqref{supp_eq:onsite_sym}, the corresponding flux-insertion operators realizes a projective representation for $g_1$ and  $g_2$ in the virtual space, so the large gauge transformation operator associated with $\hat{U}_{g_1}\hat{U}_{g_2}\hat{U}_{g_1g_2}^\dagger$ is proportional to the identity, and cannot change the symmetry charge of $\hat\rho_{q_g}$, a contradiction.

\subsection{Fermionic models with chiral-symmetry-charge flow}
We now turn to fermionic models with chiral symmetry-charge flow, satisfying $[\hat{\mathcal{F}}(\pi)]^{-1}(-1)^{\hat{Q}_\chi} \hat{\mathcal{F}}(\pi)= (-)(-1)^{\hat{Q}_\chi}$. The key observations are:
\begin{itemize}
    \item $(-1)^{\hat{Q}_\chi}\sim(-1)^{\hat{Q}}$ up to an irrelevant sign factor (for its derivation via the Jordan-Wigner transformation, see Sec.~\ref{supp_sec:stagger_fermion_JW});
    \item For a short-range entangled fermionic pure state, the $U(1)$ symmetry implies that the $\pi$-flux insertion operator commutes with the fermion parity operator~\cite{bultinck2017prb}.
\end{itemize}
As in the bosonic case, this forces the purified state to be long-range entangled.-- By contradiction, if the purified state were short-range entangled, inserting a $\pi$ $U(1)$ flux could not change the chiral symmetry charge.

\subsection{Onsite realization of symmetry on an enlarged Hilbert space \label{supp_sec:onsite_rep}}
We demonstrate that the symmetry $\hat{U}_g$ defined in Appendix~\ref{supp_sec:flux_tn}, admits an onsite representation on an enlarged Hilbert space. Specifically,  $\hat{U}_g$ is given by 
\begin{equation}
\hat{U}_{g}=\begin{array}{c}
	\begin{tikzpicture}[scale=.4, baseline={([yshift=0ex]current bounding box.center)}, thick]
		\ATensorMPO{0, 1}{$A_g$}
        \ATensorMPO{2, 1}{$A_g$}
        \draw[shift={(0,0)},dotted] (2.8,1) -- (5,1);
        \ATensorMPO{6, 1}{$A_g$}
        \draw[shift={(0,2.8)}] (-1.2,-1.8) arc(90:270:0.2);
         \draw[shift={(0,0)}] (7.2,0.6) arc(-90:90:0.2);
	\end{tikzpicture}\end{array}, 
\end{equation}
where \begin{equation}
 \begin{array}{c}
  \begin{tikzpicture}[scale=0.4, baseline={([yshift=0ex]current bounding box.center)}, thick]
      \ATensorMPO{2, 1}{$A_g$}
  \end{tikzpicture}
  \end{array}=
  \begin{array}{c}
  \begin{tikzpicture}[scale=0.4, baseline={([yshift=0ex]current bounding box.center)}, thick]
       \draw (0, -1)--(0, 1);
       \draw (-0.75, 0)--(0, 0);
       \draw (0.75, 0)--(0, 0);
       \wTensor{0, 0.5}{$R_g$}
       \xTensor{0.9, 0}{\scalebox{1}{$\alpha_g$}}

       \draw (0.5,-1.2)   node {\scriptsize $g_1$};
       \draw (0.8,2)   node {\scriptsize $g_1^\prime$};

       \draw (-1,-0.5)   node {\scriptsize $g_l$};
       \draw (1.8,-0.5)   node {\scriptsize $g_r$};
  \end{tikzpicture}
  \end{array}= \delta_{g_1^\prime, g_1g^{\scalebox{0.5}{$-1$}}}\delta_{g_1, g_l} \alpha(g_1 g_r^{-1}, g_r g^{-1}, g),
\end{equation}
and the junction corresponds to a Dirac-delta function, i.e., $\begin{array}{c} \begin{tikzpicture}[scale=0.4, baseline={([yshift=-0.4ex]current bounding box.center)}, thick]
		\draw (-0.5,1) -- (0.5,1);
		\draw (0, 0.5) -- (0,1.5);
        \draw (0,0.2)   node {\scriptsize $g_1$}; 
        \draw (0, 1.8)   node {\scriptsize $g_2$}; 
        \draw (-0.9, 1.1)   node {\scriptsize $g_3$}; 
        \draw (1, 1.1)   node {\scriptsize $g_4$}; 
	\end{tikzpicture}
	\end{array}\equiv\delta_{g_1, g_2, g_3, g_4}$.
We shall show that this symmetry operator can be represented on an enlarged Hilbert space as $\hat{U}_g= w_g^{\otimes N}\hat{P}_e$ in an enlarged Hilbert space, where $\hat{P}_e$ projects onto the original (physical) Hilbert space, i.e., 
\begin{equation}
\hat{P}_{e}=\begin{array}{c}
	\begin{tikzpicture}[scale=.4, baseline={([yshift=0ex]current bounding box.center)}, thick]
		\ATensorMPO{0, 1}{$\mathcal{A}_e$}
        \ATensorMPO{2, 1}{$\mathcal{A}_e$}
        \draw[shift={(0,0)},dotted] (2.8,1) -- (5,1);
        \ATensorMPO{6, 1}{$\mathcal{A}_e$}
        \draw[shift={(0,2.8)}] (-1.2,-1.8) arc(90:270:0.2);
         \draw[shift={(0,0)}] (7.2,0.6) arc(-90:90:0.2);
	\end{tikzpicture}\end{array}, 
\end{equation}
and 
\begin{equation}
w_g=
 \begin{array}{c}
  \begin{tikzpicture}[scale=0.4, baseline={([yshift=-1.9ex]current bounding box.center)}, thick]
       \draw (-1, -1)--(-1, 1);
       \draw (1, -1)--(1, 1);
       \xTensor{0, 0}{$\alpha_g$}
       \wTensor{-1, 0.5}{$R_g$}
       \wTensor{1, 0.5}{$R_g$}
  \end{tikzpicture}
  \end{array},\ \begin{array}{c}\begin{tikzpicture}[scale=0.4, baseline={([yshift=-1.9ex]current bounding box.center)}, thick]
       \ATensorMPO{0, 0}{$\mathcal{A}_e$}
  \end{tikzpicture}
  \end{array}=
  \begin{array}{c}
  \begin{tikzpicture}[scale=0.4, baseline={([yshift=-1.9ex]current bounding box.center)}, thick]
       \draw (-0.5, -1)--(-0.5, 1);
       \draw (0.5, -1)--(0.5, 1);
       \draw (-1.5, 0)--(-0.5, 0);
       \draw (1.5, 0)--(0.5, 0);
  \end{tikzpicture}.
  \end{array}
\end{equation}
Here, the trivalent junction represents a Dirac-delta function.

We do this by showing that $\hat{U}_g$ can be recovered from $w_g^{\otimes N}\hat{P}_e$ by removing the redundancy in the tensor-network representation. Namely, $\hat{P}_e$ has an additional symmetry 
$d_{1,2}^{\otimes N}\hat{P}_e=\hat{P}_e$ with $d_{1,2}=w_{g_1}w_{g_2}w_{g_1g_2}^\dagger$, implying that $\hat{U}_g$ lies in a fixed $d$-charge sector.
We thus eliminate this redundancy by working directly in the reduced Hilbert space in which $\hat{P}_e$ acts as the identity. To this end, we shift the unit cell of $\hat{P}_e$ by one site, so that 
   \begin{equation}
  \begin{array}{c}
\begin{tikzpicture}[scale=0.4, baseline={([yshift=-1.9ex]current bounding box.center)}, thick]
       \ArgfpTensorMPO{0, 0}{$\tilde{\mathcal{A}}_e$}
  \end{tikzpicture}
  \end{array}=
  \begin{array}{c}
  \begin{tikzpicture}[scale=0.4, baseline={([yshift=-1.9ex]current bounding box.center)}, thick]
       \draw (-0.75, -1)--(-0.75, 1);
       \draw (0.75, -1)--(0.75, 1);
       \draw (-0.75, 0)--(0.75, 0);
  \end{tikzpicture}
  \end{array}\implies 
   \begin{array}{c}
  \begin{tikzpicture}[scale=0.4, baseline={([yshift=-1.9ex]current bounding box.center)}, thick]
       \draw (0, -1)--(0, 1);
  \end{tikzpicture}
  \end{array}.
\end{equation}
Here, the implication follows because the Dirac-delta constraints at the two trivalent junctions force the physical indices to coincide. We may therefore identify them as a single site, thereby reducing the Hilbert space so that $\hat{U}_e$ acts as the identity. In this reduced basis, we have reproduced $\hat{U}_g$.

\section{Symmetry-charge flow in the $K$-matrix framework}

For illustration, we study symmetry-charge flow in continuum models within the $K$-matrix framework \cite{wen2004oxford,lu2012prb}. The Gibbs states of these models provide concrete examples that realize anomalous symmetries and illustrate algebraic relations among symmetry operators, as well as their connection to large gauge transformations.

Specifically, we consider anomalous symmetries in the following bosonic theory~\cite{wen2004oxford,lu2012prb}:
\begin{equation}\label{supp_eq:model_K_matrix}
S=\frac{1}{4\pi}\int dtdx\sum_{I, J}^{N_K}\left(K_{IJ}\,\partial_{t}\phi_{I}\partial_{x}\phi_{J}
-V_{I J}\partial_{x}\phi_{I}\partial_{x}\phi_{J}\right),
\end{equation}
where $\phi_I\simeq \phi_I +2\pi$ is a compact scalar field, $V_{IJ}$ is a positive-definite constant matrix, and $K$ is an $N_K\times N_K$ symmetric integer-valued matrix. Interpreted as an effective edge theory of symmetry-protected topological (SPT) phases, the model with $|\det K|=1$ captures the associated symmetry anomalies~\cite{lu2012prb}.
In the mixed-state setting, it also provides a concrete framework for describing the algebra of symmetry and large gauge transformations.

We review the construction of Ref.~\cite{lu2012prb}, focusing on bosonic SPTs, while fermionic ones are treated there analogously. For a unitary finite Abelian symmetry element $g\in G$, the bosonic fields transform as~\cite{lu2012prb}
\begin{equation}
\phi_I\rightarrow \sum_{J}\left(W^{g}\right)_{IJ}\phi_J+\delta\phi_I^{g},
\end{equation}
where $W^{g}\in GL(N_K,\mathbb{Z})$ and $\delta\phi_I^{g}$ is a constant shift. Accordingly, symmetry implies that,
\begin{equation}
\left(W^{g}\right)^{T}KW^{g}=K.
\end{equation}
A key simplification is that a $2\times 2$ $K$ matrix provides a minimal building block for the topological structure, while more general $K$ matrices can be obtained by taking direct sums of these $2\times 2$ blocks. More importantly, Ref.~\cite{lu2012prb} shows that the $2\times 2$ $K$ matrix can be brought to the standard form,
\begin{equation}\label{supp_eq:canonical_K}
K=\sigma^{x}=\begin{pmatrix}
0 & 1\\
1 & 0
\end{pmatrix},
\end{equation}
while different forms can be related via a relabeling of the $\phi$ fields, i.e., $\phi \rightarrow  X \phi$, $\det X\neq 0$.
Distinct phases with $\mathbb{Z}_{\mathcal{N}_g}$ symmetry are then labeled by an integer $q\in\{0, 1,2,\ldots,\mathcal{N}_g-1\}$ through the symmetry action
\begin{equation}\label{supp_eq:sym_kmatrix}
\phi_I\rightarrow \phi_I +\frac{2\pi}{\mathcal{N}_g}u^g_I,\  \text{and}\ (u^g)^T=(1, -q),
\end{equation} 
where $W^g=\mathbb{I}$.
For the general finite Abelian case, it follows similarly because any finite Abelian group is isomorphic to $\mathbb{Z}_{\mathcal{N}_1}\times \mathbb{Z}_{\mathcal{N}_2}\dots$, with different $u^g$. However, this construction captures the anomaly for each $\mathbb{Z}_{\mathcal{N}_I}$ factor and for each pair $\mathbb{Z}_{\mathcal{N}_I}\times \mathbb{Z}_{\mathcal{N}_J}$, while the triple-factor case is less clear. We therefore delegate this case to the tensor-network models (Appendix~\ref{supp_sec:flux_tn}).

With these preparations, we now turn to the operator formulation in order to construct the flux-insertion operator [see Eqs.~(\ref{supp_eq:flux_regularization}, \ref{supp_eq:flux_insertion}) for results], and to clarify its relation to large gauge transformations. The key input is the equal-time commutation relation (which follows directly from the action above),
\begin{equation}
\sum_{J}\Big[K_{IJ}\,\partial_{x}\hat{\phi}_{J}(x),\ \partial_{y}\hat{\phi}_{K}(y)\Big]
=2\pi i\,\delta_{IK}\,\partial_{x}\delta(x-y).
\end{equation}
In turn, the operator for the symmetry transformation Eq.~\eqref{supp_eq:sym_kmatrix} is 
\begin{equation}
\hat{U}_{g}=e^{i\frac{1}{\mathcal{N}_g}(u^g)^T K \left(\oint dx\partial_{x}\hat{\phi}\right)} \implies \hat{U}_g\, \hat\phi\, \hat{U}_g^\dagger =\hat\phi+ \frac{2\pi}{\mathcal{N}_g}u^g.
\end{equation}
Consequently, we can infer the flux-insertion operator. For open boundary conditions on the interval
$x\in[x_a,x_b]$, we obtain
\begin{eqnarray}\label{supp_eq:flux_regularization}
&&\left[\hat{U}_g^{(\mathrm{o})}\right]^{\mathcal{N}_g}\nonumber\\
&=&\lim_{\epsilon\to 0^+}
\left[e^{iq\hat{\phi}_1(x_a)}\,e^{-i\hat{\phi}_2(x_a-\epsilon)}\right]
\left[e^{-iq\hat{\phi}_1(x_b)}\,e^{i\hat{\phi}_2(x_b+\epsilon)}\right]\nonumber\\
&=&
e^{iq\hat{\phi}_1(x_a)}\times e^{-iq\hat{\phi}_1(x_b)} |_{x\in \left[x_a, x_b\right]},
\end{eqnarray}
where $\epsilon$ is a short-distance regularization~\cite{else2014prb}, while the symmetry-charge flow is independent of whether $\phi_1$ or $\phi_2$ is regularized. Specifically, the flux-insertion operator, associated with $(\hat{U}_g)^{\mathcal{N}_g}$ that generates a large gauge transformation, is
\begin{equation}
\hat{\mathcal{F}}_{\mathcal{L}}=e^{iq\hat \phi_1(x_a)}\equiv e^{-i (u^g)^T KP_{\text{reg}}\hat\phi(x_a)} ,
\end{equation}
with $P_{\text{reg}}=\text{diag}(1, 0)$ from the regularization. It satisfies
\begin{equation}\label{supp_eq:flux_insertion}
\hat{\mathcal{F}}_{\mathcal{L}}^{-1}\, \hat{U}_g\,  \hat{\mathcal{F}}_{\mathcal{L}}=e^{i\Delta \mathcal{S}_g}\, \hat{U}_g,
\end{equation}
with 
\begin{equation}
\Delta \mathcal{S}_g = -\frac{2\pi}{\mathcal{N}_g} [(u^g)^T P_{\text{reg}}K u^g]=\frac{2\pi q}{\mathcal{N}_g},
\end{equation}
which indicates the presence of symmetry-charge flow. For a generic Abelian symmetry, the phase factor $\Delta\mathcal{S}_g$ [Eq.~\eqref{supp_eq:flux_insertion}] contains mixed terms in which one $u^g$ is replaced by $u^{g_1}$ or $u^{g_2}$ associated with the other $\mathbb{Z}_{\mathcal{N}_{1,2}}$ groups. Finally, we remark that even starting from a $K$ with $K\neq\sigma^{x}$, one can choose a basis in which the symmetry-charge flow remains manifest, via the transformation $\phi \rightarrow X \phi$, with $\det X\neq 0$. We also confirm this in the tensor-network models in Sec.~\ref{supp_sec:standard_unitary}.

The connection between $\hat{\mathcal{F}}_{\mathcal{L}}$ and a large gauge transformation is made explicit by the following formula,
\begin{equation}
\oint \partial_x \hat{\phi}_I \;\rightarrow\; \oint \partial_x \hat{\phi}_I + 2\pi q\,\delta_{I,2}\, ,
\end{equation}
obtained by using
\begin{equation}
\hat{\mathcal{F}}_{\mathcal{L}}\,\partial_x \hat\phi_I(x)\,\hat{\mathcal{F}}_{\mathcal{L}}^{-1}
= \partial_x \hat\phi_I(x)+ 2\pi q\,\delta_{I,2}\,\delta(x-x_a).
\end{equation}
Here, $q\in \{0, 1, 2,\dots \mathcal{N}_g-1\}$, thus corresponding to a flux shift of $2\pi q$.

\section{Chiral anomaly in the fermion models}

We provide details of Eq.~\eqref{eq:chiralA_finiteT} in the main text,
\begin{equation}
\partial_\mu \langle \hat j^\mu_\chi\rangle = \mathcal{C}(\beta)\epsilon^{\mu\nu}\partial_\mu A_\nu,
\end{equation}
using two complementary calculations that emphasize different aspects of the anomaly. The linear-response calculation in Sec.~\ref{supp_sec:perturbative_chiral} is conceptually straightforward, but requires a choice of gauge. By contrast, the calculation in Sec.~\ref{supp_sec:anomaly_nonperturbative} embeds the fermionic Hamiltonian into a theory with a dynamical $U(1)$ gauge field, which avoids an explicit gauge fixing and provides a manifestly gauge-invariant anomalous response.

\subsection{Chiral anomaly from linear response \label{supp_sec:perturbative_chiral}}
To calculate the anomaly, we first recall that the Hamiltonian is
\begin{equation}
\hat{H}_{\text{sf}}=-i\sum_{j=1}^{N}\left(\hat{\psi}_{j}^{\dagger}\hat{\psi}_{j+1}-\hat{\psi}_{j+1}^{\dagger}\hat{\psi}_{j}\right),
\end{equation}
with $U\left(1\right)$ and $U_{\chi}\left(1\right)$ symmetries,
i.e., 
\begin{equation}
\begin{cases}
\hat{Q}=\sum_{j}\hat{j}_{j}^{0}=\sum_{j}\hat{\psi}_{j}^{\dagger}\hat{\psi}_{j}\\
\hat{Q}_{\chi}=\sum_{j}\hat{j}_{\chi, j}^{0}=\frac{1}{2}\sum_{j=1}^{N}\left[\left(\hat{\psi}_{j}+\hat{\psi}_{j}^{\dagger}\right)\left(\hat{\psi}_{j+1}-\hat{\psi}_{j+1}^{\dagger}\right)+1\right]
\end{cases}.
\end{equation}

We compute the response current in the gauge $A_{\mu}=\left(A_{0},\ 0\right)$, by expanding perturbatively in $A_0$.
Specifically, to leading order in $A_0$, the time evolution of the chiral charge density is, 
\begin{eqnarray}\label{supp_eq:time_evo_j_chi}
\partial_{t}\langle\hat{j}_{\chi,\ i}^{0}\rangle & = & i\langle\left[\hat{H}_{\text{sf}}+\delta\hat{H}_{\text{sf}},\ \hat{j}_{\chi,\ i}^{0}\right]\rangle\nonumber \\
 & = & i\sum_{j}\langle\left[\hat{j}_{j}^{0},\ \hat{j}_{\chi,\ i}^{0}\right]\rangle A_{0,\, j}.
\end{eqnarray}
In the first line, the perturbed Hamiltonian is 
\begin{equation}
\hat{H}_{\text{sf}}\left(A_{0}\right)=\hat{H}_{\text{sf}}+\delta\hat{H}_{\text{sf}}\left(A_{0}\right),
\end{equation}
with 
\begin{equation}
\begin{cases}
\hat{H}_{\text{sf}}=-i\sum_{j=1}^{N}\left(\hat{\psi}_{j}^{\dagger}\hat{\psi}_{j+1}-\hat{\psi}_{j+1}^{\dagger}\hat{\psi}_{j}\right)\\
\delta\hat{H}_{\text{sf}}\left(A_{0,i}\right)=\sum_{j}A_{0, j}\,\hat{j}_{j}^{0},\ \ \hat{j}_{j}^{0}=\hat{\psi}_{j}^{\dagger}\hat{\psi}_{j}
\end{cases}.
\end{equation}
The second line uses $\langle\left[\hat{H}_{\text{sf}},\ \hat{j}_{\chi,\ i}^{0}\right]\rangle=0$, implied by translation invariance and chiral symmetry, i.e.,
\begin{equation}
\langle\left[\hat{H}_{\text{sf}},\ \hat{j}_{\chi,\ i}^{0}\right]\rangle=\frac{1}{N}\langle\left[\hat{H}_{\text{sf}},\ \hat{Q}_{\chi}\right]\rangle=0.
\end{equation}

Crucially, the commutator in the time-evolution equation Eq.~\eqref{supp_eq:time_evo_j_chi} encodes the anomaly, and reproduces the Schwinger term in the zero-temperature, continuum limit~\cite{chatterjee2025prl},
\begin{eqnarray}\label{supp_eq:lattice_schwinger}
\left[\hat{j}_{i}^{0},\ \hat{j}_{\chi,\ j}^{0}\right] & = & \overset{\text{Schwinger term}}{\underline{\left(\delta_{i,\ j}-\delta_{i,\ j+1}\right)\frac{1}{2}\left(\hat{\psi}_{j}^{\dagger}\hat{\psi}_{j+1}-\hat{\psi}_{j+1}^{\dagger}\hat{\psi}_{j}\right)}}\nonumber \\
 &  & +\left(\delta_{i,\ j}+\delta_{i,\ j+1}\right)\frac{1}{2}\left(\hat{\psi}_{j+1}^{\dagger}\hat{\psi}_{j}^{\dagger}+\hat{\psi}_{j+1}\hat{\psi}_{j}\right).\nonumber \\
\end{eqnarray}
In particular, in the zero-temperature continuum limit, the first term reduces to the Schwinger term,
\begin{equation}
\left[\hat{j}^{0}(x),\ \hat{j}_{\chi}^{0}(y)\right]_G\sim-i\frac{1}{\pi}\partial_{x}\delta\left(x-y\right),
\end{equation}
where the subscript $G$ denotes the zero-temperature limit. Here, we write $j^0_i$ and $j^0_{\chi,j}$ as $j^0(x)$ and $j^0_{\chi}(y)$, respectively, and the coefficient $-\frac{1}{\pi}$ arises from the ground-state expectation value.
The second term in Eq.~\eqref{supp_eq:lattice_schwinger} breaks $U\left(1\right)$ symmetry, and
vanishes in density-matrix expectation values. 

Putting these results together, the chiral anomaly takes the form
\begin{equation}
\partial_{t}\langle\hat{j}_{\chi,\ i}^{0}\rangle=-\mathcal{C}\left(\beta\right)\partial_{x}A_{0}.
\end{equation}
The coefficient is, 
\begin{eqnarray*}
\mathcal{C}\left(\beta\right) & = & -i\frac{1}{2}\frac{1}{N}\sum_{j}\langle\left(\hat{\psi}_{j}^{\dagger}\hat{\psi}_{j+1}-\hat{\psi}_{j+1}^{\dagger}\hat{\psi}_{j}\right)\rangle\\
 & = & \frac{1}{N}\sum_{q}\sin q\langle\hat{\psi}_{q}^{\dagger}\hat{\psi}_{q}\rangle.
\end{eqnarray*}
In the first equality, we have used translation symmetry to replace the
local expectation value by its spatial average. In the second, we
have performed a Fourier transform, i.e., $\hat{\psi}_{j}=\frac{1}{\sqrt{N}}\sum_{q}e^{iqx_{j}}\hat{\psi}_{q}$, 
which in the continuum limit becomes, 
\begin{eqnarray}\label{supp_eq:chiral_coeff_perturbative}
\mathcal{C}\left(\beta\right) & = & \int\frac{dq}{2\pi}\sin q\frac{e^{-2\beta\sin q}}{1+e^{-2\beta\sin q}}\nonumber \\
 & = & \frac{1}{2}\int\frac{dq}{2\pi}\sin q\left(\frac{2e^{-2\beta\sin q}}{1+e^{-2\beta\sin q}}-1\right)\nonumber \\
 & = & \frac{1}{2}\int\frac{dq}{2\pi}\sin q\left(\frac{e^{-2\beta\sin q}-1}{1+e^{-2\beta\sin q}}\right)\nonumber \\
 & = & -\frac{1}{2}\int\frac{dq}{2\pi}\sin q\tanh\left(\beta q\right).
\end{eqnarray}

\subsection{Chiral anomaly from a gauge-invariant  calculation \label{supp_sec:anomaly_nonperturbative}}
We also present an alternative calculation, and
reproduce the result above. 

To this end, we first consider a
modified chiral charge~\cite{chatterjee2025prl},
\begin{equation}
\hat{\tilde{Q}}_{\chi}=\frac{1}{2}\sum_{j}\left(\hat{\psi}_{j}^{\dagger}\hat{\psi}_{j+1}-\hat{\psi}_{j}\hat{\psi}_{j+1}^{\dagger}+1\right),
\end{equation}
which is invariant under the global $U\left(1\right)$ transformation.
It differs from $\hat{Q}_{\chi}$ by 
\begin{equation}
\hat{Q}_{\chi}-\hat{\tilde{Q}}_{\chi}=\frac{1}{2}\sum_{j=1}^{N}\left(\hat{\psi}_{j}\hat{\psi}_{j+1}-\hat{\psi}_{j}^{\dagger}\hat{\psi}_{j+1}^{\dagger}\right),
\end{equation}
a term that breaks global $U\left(1\right)$ to $\mathbb{Z}_{2}$.
Consequently, it vanishes in the expectation value $\langle\dots\rangle=\frac{\text{Tr}e^{-\beta\hat{H}\left(A\right)}\left(\dots\right)}{\text{Tr}e^{-\beta\hat{H}\left(A\right)}}$,
and therefore does not affect the anomaly equation.

Below, we complete the calculation by first introducing a dynamical $U(1)$ gauge field, from which a constant external electric field emerges upon projecting back to a fixed electric-field sector. 

\subsubsection{Model with $U\left(1\right)$ gauge fields }

\paragraph{Gauged Hamiltonian.--} We start with the following gauged Hamiltonian~\cite{banks1976prd, carroll1976prd,hamer1997prd, dempsey2022prr}, 
\begin{eqnarray}
\hat{H}\left(A\right) & = & \sum_{j}\hat{h}_{j+\frac{1}{2}}\left(A\right)+\sum_{j}\hat{E}_{j+\frac{1}{2}}^{2},
\end{eqnarray}
with 
\begin{equation}
\hat{h}_{j+\frac{1}{2}}\left(A\right)=-i\left(\hat{\psi}_{j+1}^{\dagger}\hat{U}_{j+\frac{1}{2}}\hat{\psi}_{j}+\hat{\psi}_{j}\hat{U}_{j+\frac{1}{2}}^{\dagger}\hat{\psi}_{j+1}^{\dagger}\right).
\end{equation}
Here, $\hat{U}_{j+\frac{1}{2}}=e^{-i\hat{A}_{j+1,\ j}}$ is the Wilson
line associated with the gauge field $\hat{A}_{j+1,\ j}$. $\hat{E}_{j+\frac{1}{2}}$
is the conjugate electric field, obeying 
\begin{equation}
\left[\hat{E}_{j+\frac{1}{2}},\ \hat{A}_{i+1,\ i}\right]=-i\delta_{ij}\implies\left[\hat{E}_{j+\frac{1}{2}},\ \hat{U}_{i+1,\ i}\right]=-\hat{U}_{i+\frac{1}{2}}\delta_{ij}.
\end{equation}
Additionally, for notational simplicity, we have absorbed the matter-gauge-field coupling constant into the definition of the $U(1)$ gauge field, so that the perturbative expansion below is performed directly in $\hat{A}$.

\paragraph{Gauged chiral symmetry charge.--} Similarly, we can write $\hat{\tilde{Q}}_{\chi}$ in a gauge-invariant
form, i.e., 
\begin{equation}
\hat{\tilde{Q}}_{\chi}\left(A\right)=\frac{1}{2}\sum_{j}\left(\hat{\psi}_{j}^{\dagger}\hat{U}_{j+\frac{1}{2}}\hat{\psi}_{j+1}-\hat{\psi}_{j}\hat{U}_{j+\frac{1}{2}}^{\dagger}\hat{\psi}_{j+1}^{\dagger}+1\right).
\end{equation}
Its time evolution then follows from the Heisenberg equation,
\begin{eqnarray}
\partial_{t}\hat{\tilde{Q}}_{\chi} & = & -i\left[\hat{\tilde{Q}}_{\chi},\ \hat{H}\left(A\right)\right]\nonumber \\
 & = & -i\left[\hat{\tilde{Q}}_{\chi},\ \sum_{j}\hat{E}_{j+\frac{1}{2}}^{2}\right]\nonumber \\
 & = & \frac{1}{2}\sum_{j}\left\{ \hat{E}_{j+\frac{1}{2}},\ \hat{h}_{j}\left(A\right)\right\} .
\end{eqnarray}

\paragraph{External electric field from restricting the gauge-field Hilbert
space.--}
To compute the chiral anomaly induced by an external electric field,
we restrict to the sector with a fixed electric field, i.e., 
\begin{equation}
|\Psi_{n}\left(E\right)\rangle=|E\rangle\otimes|\psi_{n}\rangle,\ \hat{E}_{i+\frac{1}{2}}|E\rangle=E_{i+\frac{1}{2}}|E\rangle,
\end{equation}
and 
\begin{equation}
\langle\Psi_{n}\left(E\right)|\hat{H}\left(A\right)|\Psi_{n}\left(E\right)\rangle=\mathcal{E}_{n}.
\end{equation}

\subsubsection{Anomaly equation }
After these preparations, we can compute the anomaly
equation associated with $\partial_{t}\hat{\tilde{Q}}_{\chi}$, 
\begin{eqnarray}
\langle\partial_{t}\hat{\tilde{Q}}_{\chi}\rangle & = & \frac{1}{2}\sum_{j}\langle\left\{ \hat{E}_{j+\frac{1}{2}},\ \hat{h}_{j+\frac{1}{2}}\left(A\right)\right\} \rangle\nonumber \\
 & = & \sum_{j}E_{j+\frac{1}{2}}\mathcal{C}(\beta)+\mathcal{O}(\hat{A}^2),
\end{eqnarray}
where we have retained only terms linear in the gauge field.
Here, the coefficient $\mathcal{C}\left(\beta\right)$ reproduces the linear-response result obtained above. Specifically, we have
\begin{equation}
\mathcal{C}\left(\beta\right)=\frac{1}{N}\sum_{j}\frac{\text{Tr}\left[e^{-\beta\hat{H}\left(A\right)}\hat{h}_{j+\frac{1}{2}}\left(A\right)\right]}{\text{Tr}\left[e^{-\beta\hat{H}\left(A\right)}\right]}|_{A=0}.
\end{equation}
This follows from translation symmetry, 
\begin{eqnarray}
&&\text{Tr}\left[e^{-\beta\hat{H}\left(A\right)}\hat{h}_{j+\frac{1}{2}}\left(A\right)\right]|_{A=0}\nonumber\\
&=&\frac{1}{N}\text{Tr}\left[e^{-\beta\hat{H}\left(A\right)}\hat{H}\left(A\right)\right]|_{A=0}.
\end{eqnarray}
In particular, $\mathcal{C}(\beta)$ is the thermal
expectation value of the local energy density, and in the continuum
limit, it becomes, 
\begin{eqnarray}
\mathcal{C}\left(\beta\right) & = & -\frac{1}{2}\frac{1}{N}\partial_{\beta}\ln\text{Tr}e^{-\beta\hat{H}}\nonumber \\
 & = & -\frac{1}{2}\partial_{\beta}\frac{1}{N}\sum_{q}\ln\left(1+e^{-2\beta\sin q}\right)\nonumber \\
 & = & \int\frac{dq}{2\pi}\sin q\frac{e^{-2\beta\sin q}}{1+e^{-2\beta\sin q}}\nonumber \\
 & = & \int\frac{dq}{2\pi}\sin q\frac{1}{2}\left[\frac{2e^{-2\beta\sin q}}{1+e^{-2\beta\sin q}}-1+1\right]\nonumber \\
 & = & -\frac{1}{2}\int\frac{dq}{2\pi}\sin q\tanh\left(\beta\sin q\right),
\end{eqnarray}
in agreement with the perturbative results.

\subsection{Symmetry-charge flow from zero-temperature Schwinger term \label{supp_sec:Symmetry_charge_SchwingerTerm}}
We provide details of the chiral symmetry-charge flow implied by the chiral anomaly, by deriving
\begin{equation}\label{supp_eq:distribution_ground_state1}
\hat{\mathcal{F}}(\phi)^{-1}\hat{Q}_\chi\hat{\mathcal{F}}(\phi)=\hat{Q}_\chi +\frac{1}{\pi}\phi,
\end{equation}
where $\hat{Q}_\chi$ is the chiral symmetry generator and its eigenvalues are denoted by $q_\chi$. $\hat{\mathcal{F}}(\phi)$ is the flux-insertion operator for the $U(1)$ flux $\phi$.
Consequently, Eq.~\eqref{supp_eq:distribution_ground_state1} implies that under a large gauge transformation $\phi\rightarrow \phi+\Delta\phi$ (with $\Delta\phi=2\pi$), the weight of the $q_\chi$ sector in the ground-state density matrix transforms as
\begin{equation}
p_{q_{\chi}}\left(\phi+\Delta \phi\right)=p_{q_{\chi}-\frac{1}{\pi}\Delta\phi}\left(\phi\right),\label{supp_eq:distribution_ground_state}
\end{equation}
where the ground-state density matrix, $\hat{\rho}_{G}\left(\phi_{x}\right)$, decomposes in the $\hat{Q}_\chi$ eigenspace as,
\begin{equation}
\begin{cases}
\hat{\rho}_{G}\left(\phi\right)\equiv\sum_{q_{\chi}}p_{q_{\chi}}\left(\phi\right)\hat{\rho}_{q_{\chi}}\left(\phi\right)\\
\hat{\rho}_{G,\ q_{\chi}}\left(\phi\right)\equiv\frac{\hat{P}_{q_{\chi}}\hat{\rho}_{G}\left(\phi\right)\hat{P}_{q_{\chi}}}{\text{Tr}\left[\hat{P}_{q_{\chi}}\hat{\rho}_{G}\left(\phi\right)\hat{P}_{q_{\chi}}\right]}
\end{cases},
\end{equation}
and $p_{q_\chi}\equiv \text{Tr}[\hat\rho(\phi)\hat{P}_{q_\chi}]$ with $\hat{P}_{q_{\chi}}$ the projector onto the $q_{\chi}$-eigenspace
of $\hat{Q}_{\chi}$. 

The key ingredient for Eq.~\eqref{supp_eq:distribution_ground_state1} is the anomaly equation in the zero-temperature limit, 
\begin{equation}
\partial_{\mu}\langle\hat{j}_{\chi}^{\mu}\rangle_{G}=-\frac{1}{\pi}\epsilon^{\mu\nu}\partial_{\mu}A_{\nu},
\end{equation}
or equivalently, from the Schwinger term~\cite{fradkin2021princeton},
\begin{equation}
\left[\hat{j}^{0}(x),\ \hat{j}_{\chi}^{0}(y)\right]=-i\frac{1}{\pi}\partial_{x}\delta\left(x-y\right).
\end{equation}
Specifically, on an
infinite line, inserting a $U(1)$ flux $\phi$ at site
$x_{0}$, is implemented by
\begin{equation}
\hat{\mathcal{F}}\left(\phi\right)\equiv e^{i\phi\hat{Q}_{[x_{0}, \infty)}},
\end{equation}
where $\hat{Q}_{[x_{0}, \infty)}$ is $\hat{Q}$ restricted to $[x_{0}, \infty)$.
The Schwinger term then implies 
\begin{equation}
\left[\hat{Q}_{[x_{0}, \infty)},\ \hat{j}_{\chi}^{0}(y)\right]=i\frac{1}{\pi}\delta\left(x_{0}-y\right),
\end{equation}
and thus 
\begin{equation}
\left[\hat{\mathcal{F}}\left(\phi\right)\right]^{-1}\hat{j}_{\chi}^{0}(y)\hat{\mathcal{F}}\left(\phi\right)=\hat{j}_{\chi,}^{0}(y)+\frac{1}{\pi}\phi\delta\left(x_{0}-y\right),\label{supp_eq:flux_insertion_0T}
\end{equation}
which reproduces Eq.~\eqref{supp_eq:distribution_ground_state1}. 
Finally, Eq.~\eqref{supp_eq:distribution_ground_state} follows by applying the identities below to the definition of $p_{q_\chi}$,
\begin{equation}
\begin{cases}
\hat{\rho}_{G}\left(\phi+\Delta \phi\right)=\hat{\mathcal{F}}\left(\Delta\phi\right)\hat{\rho}_{G}\left(\phi\right)[\hat{\mathcal{F}}\left(\phi\right)]^{-1}\\
\left[\hat{\mathcal{F}}\left(\Delta\phi\right)\right]^{-1}\hat{P}_{q_{\chi}}\hat{\mathcal{F}}\left(\Delta\phi\right)=\hat{P}_{q_{\chi}-\frac{1}{\pi}\Delta\phi}
\end{cases}.
\end{equation}

\section{Derivation of Eq.~(\ref{eq:flux_chi_sym}) \label{supp_sec:stagger_fermion_JW}}
We provide details for Eq.~\eqref{eq:flux_chi_sym} in the main text, reproduced here for convenience,
\begin{equation}
\left[\hat{\mathcal{F}}\left(\pi\right)\right]^{-1}e^{i\pi\hat{Q}_{\chi}}\hat{\mathcal{F}}\left(\pi\right)=\left(-1\right)e^{i\pi\hat{Q}_{\chi}},
\end{equation}
where $\hat{\mathcal F}(\pi)$ implements the $\pi$ $U(1)$-flux insertion on an \textit{infinite} chain, and $\hat{Q}_\chi$ is the chiral symmetry charge.
We shall establish the flux-twisted chiral-parity relation on an
$N$-site periodic chain using the quotient construction of
Appendix~\ref{supp_sec:flux_insertion}. The infinite-chain relation is
then obtained in the $N\to\infty$ limit.

The derivation uses the Jordan-Wigner transformation (see Sec.~\ref{supp_sec:JW_trans}) to map the system to a spin representation, and more importantly to simplify the gauged operator. Namely, the spin representations of $\hat{Q}$ and $\hat{Q}_\chi$ on an $N$-site periodic chain are
\begin{equation}
\hat{Q}=\frac{1}{2}\sum_i^N(1+\sigma^z_i),
\end{equation}
and 
\begin{equation}
\hat{Q}_\chi\equiv \sum_i^N \hat{j}^0_{\chi, i} =\sum_{i}^N\frac{1}{2}\left[1+\left(-1\right)^{\left(1+\hat{Q}\right)\delta_{i,\ N}}\sigma_{i+1}^{x}\sigma_{i}^{x}\right],
\end{equation}
where $\left(-1\right)^{\left(1+\hat{Q}\right)\delta_{i,\ N}}$ encodes the boundary contribution from the fermion-parity string, and $(-1)^{\hat{Q}_\chi}$ equals $(-1)^{\hat{Q}}$ up to an overall scalar factor, i.e., 
\begin{equation}\label{supp_eq:fermin_parity_chiral}
(-1)^{\hat{Q}_\chi}=(-1)^{N+1}(-1)^{\hat{Q}}.
\end{equation}

We then obtain Eq.~\eqref{eq:flux_chi_sym} via the quotient construction in Appendix~\ref{supp_sec:flux_insertion}: Starting from an infinite parent chain, we insert a $\pi$ flux at each $x_0+mN$ $(m\in\mathbb{Z})$, corresponding to one flux per unit cell. In particular, a $\pi$ flux through the link at $x_0$ flips the sign of the associated term:
\begin{equation}
\hat{j}^0_{\chi, x_0}\rightarrow \frac{1}{2}\big[1+\underline{(-1)}\sigma^x_{x_0+1}\sigma^x_{x_0}\big].
\end{equation}
Repeating the insertion at each $x_0+mN$ produces one $\pi$ flux per unit cell, and contributes an overall factor of $(-1)$ to the gauged chiral-parity operator on the $N$-site ring. This follows by simplifying the gauged operator with the identity,
\begin{equation}
\label{supp_eq:Z2gauge_chi}
(-1)^{\hat{Q}_\chi(\mathcal{A})}
=
e^{i\sum_i\mathcal{A}_{i+1,i}}
(-1)^{\hat{Q}_\chi(0)},
\end{equation}
where $(-1)^{\hat{Q}_\chi(\mathcal{A})}$ denotes the chiral-charge parity in the presence of a $\mathbb{Z}_2$ gauge field $\mathcal{A}$, with
$e^{i\mathcal{A}_{i+1,i}}=\pm1$. This identity follows directly from
\begin{equation}
e^{i\frac{\pi}{2}
\left(
1+e^{i\mathcal{A}_{i+1,i}}
\sigma^x_{i+1}\sigma^x_i
\right)}
=
-\,e^{i\mathcal{A}_{i+1,i}}
\sigma^x_{i+1}\sigma^x_i.
\end{equation}
Thus, Eq.~\eqref{supp_eq:Z2gauge_chi} contributes the overall factor of $(-1)$ on the $N$-site ring. Taking $N\to\infty$ recovers Eq.~\eqref{eq:flux_chi_sym}.

\subsubsection{Jordan--Wigner transformation convention \label{supp_sec:JW_trans}}
Here we specify our Jordan-Wigner transformation convention, \begin{equation}
\begin{cases}
\hat{\psi}_{i}^{\dagger}=\prod_{l=i+1}^{N}\left(-\sigma_{l}^{z}\right)\sigma_{i}^{+}\\
\hat{\psi}_{i}=\prod_{l=i+1}^{N}\left(-\sigma_{l}^{z}\right)\sigma_{i}^{-}
\end{cases},
\end{equation}
with $\sigma^{x, y, z}$ for the Pauli matrices and 
\begin{equation}
\sigma^{+}\equiv\left(\begin{array}{cc}
0 & 1\\
0 & 0
\end{array}\right),\ \sigma^{-}\equiv\left(\begin{array}{cc}
0 & 0\\
1 & 0
\end{array}\right).
\end{equation}
In turn, we find the following identities for fermion bilinears
\begin{equation}
\begin{cases}
\left(-1\right)^{\left(1+\hat{Q}\right)\delta_{i,\ N}}\hat{\psi}_{i+1}^{\dagger}\hat{\psi}_{i}=\sigma_{i+1}^{+}\sigma_{i}^{-}\\
\left(-1\right)^{\left(1+\hat{Q}\right)\delta_{i,\ N}}\hat{\psi}_{i+1}\hat{\psi}_{i}=-\sigma_{i+1}^{-}\sigma_{i}^{-}\\
\left(-1\right)^{\left(1+\hat{Q}\right)\delta_{i,\ N}}\hat{\psi}_{i+1}^{\dagger}\hat{\psi}_{i}^{\dagger}=\sigma_{i+1}^{+}\sigma_{i}^{+}\\
\left(-1\right)^{\left(1+\hat{Q}\right)\delta_{i,\ N}}\hat{\psi}_{i+1}\hat{\psi}_{i}^{\dagger}=-\sigma_{i+1}^{-}\sigma_{i}^{+}
\end{cases},
\end{equation}
using the following identities
\begin{equation}
\begin{cases}
\left\{ -\sigma^{z},\ \sigma^{\pm}\right\} =0\\
-\sigma^{z}\sigma^{\pm}=\mp\sigma^{\pm}\\
\sigma^{\pm}\left(-\sigma^{z}\right)=\pm\sigma^{\pm}
\end{cases}.
\end{equation}

\section{Symmetries in the reduced density matrix of a two dimensional symmetry-protected topological state \label{supp_sec:spt_mpo}}
We study the reduced density matrix of a two-dimensional symmetry-protected topological (SPT) ground state in a tensor-network representation. In particular, the result here applies equally to Abelian and non-Abelian symmetries.

\subsection{Reduced density matrix}
The reduced density matrix of a two-dimensional SPT state $\hat\rho$ can be represented as,
\begin{equation}\label{supp_eq:psi_purification}
\hat\rho =\text{Tr}_a|\Psi\rangle\langle \Psi|,
\end{equation}
where $\text{Tr}_a$ traces out the ancilla. $|\Psi\rangle$ is a pure state, with a matrix-product-state representation, 
\begin{equation}
|\Psi\rangle=
	\begin{tikzpicture}[scale=.4, baseline={([yshift=0.3ex]current bounding box.center)}, thick]
		\ATensorLPDO{0.3, 0}{$L$}
        
        \ATensorLPDO{5.7, 0}{$L$}

        \draw[shift={(0,0.8)}] (-1.2,-0.8) arc(90:270:0.2);
         \draw[shift={(0,0)}] (7.2,-0.4) arc(-90:90:0.2);
		\draw [decorate,
    	decoration = {calligraphic brace,mirror}] (-0.5,-0.8) --  (6.8,-0.8);
		\draw (2.7,-1.5) node {\scriptsize $N$};  
        \draw[shift={(0,0)},dotted] (2,0) -- (4,0);
	\end{tikzpicture}\ ,\label{supp_eq:LPDO_rep}
\end{equation}
which is known as the locally purified density-operator (LPDO) representation of the density matrix $\hat\rho$~\cite{verstrate2004prl}.
The $L$ tensor is then obtained from the object below (see Sec.~\ref{supp_sec:rdm_lpdo} below for a derivation), 
\begin{eqnarray}\label{supp_eq:peps_lpdo}
\begin{tikzpicture}[scale=.4, baseline={([yshift=-0.6ex]current bounding box.center)}, thick]
     \SPTbdTensorLPDO{0, 0}{$g_1 $}{$g_2$}{$g_4$}{$g_3$}
  \end{tikzpicture}=\frac{\alpha(g_1g_2^{-1}, g_2g_4^{-1}, g_4)}{\alpha(g_1g_3^{-1}, g_3g_4^{-1}, g_4)}\nonumber\\
\times|g_1, g_2)|g_2, g_4\rangle \ \langle g_1, g_3|\, (g_3, g_4|, \nonumber\\
\end{eqnarray}
where the edge arrow points from the larger to the smaller subscript, and the virtual index is denoted by $|\dots)$ or $(\dots|$. The black/red line is the physical/ancilla index, written as $|\dots\rangle$ or $\langle \dots|$; the connected lines have the same group indices. Accordingly, $|\Psi\rangle$ ($\begin{tikzpicture}[scale=.4, baseline={([yshift=-1.2ex]current bounding box.center)}, thick]
     \ATensorLPDO{0, 0}{$L$}
  \end{tikzpicture}$) is obtained by identifying the single red (black) line with the corresponding red (black) double lines in Eq.~\eqref{supp_eq:peps_lpdo}, and by flipping $\langle g_1, g_3|$ to a ket. This identification is assumed in the graphical notation unless stated otherwise.
  
Additionally, the triangle tensor in Eq.~\eqref{supp_eq:peps_lpdo} is derived from the PEPS tensor for two-dimensional SPTs on a directed triangular lattice~\cite{willamson2016prb},

\begin{eqnarray}\label{supp_eq:peps_triangle_bl}
\begin{tikzpicture}[scale=.4, baseline={([yshift=-1.2ex]current bounding box.center)}, thick]
    \SPTbulkTensorTri{0, 0}{$g_1 $}{$g_2$}{$g_4$}
  \end{tikzpicture}&=&\left[\alpha(g_1g_2^{-1}, g_2g_4^{-1}, g_4)\right]^{\triangle_{1,2,4}}\nonumber\\
  &&\times|g_1, g_2)\, |g_2, g_4\rangle \ (g_1, g_4|,
\end{eqnarray}
where $\triangle_{1,2,4}= +1$ for the orientation of the triangle relative to the two-dimensional lattice, as the ordering $g_1, g_2, g_4$ matches the lattice orientation. Meanwhile, $\triangle_{1,3,4}= -1$ for the following tensor,
\begin{eqnarray}\label{supp_eq:peps_triangle_ur}
\begin{tikzpicture}[scale=.4, baseline={([yshift=-1.2ex]current bounding box.center)}, thick]
    \SPTbulkTensorTriInv{0, 0}{$g_4 $}{$g_1$}{$g_3$}
  \end{tikzpicture}&=&\left[\alpha(g_1g_3^{-1}, g_3g_4^{-1}, g_4)\right]^{\triangle_{1,3,4}}\nonumber\\
  &&\times |g_1, g_4)\ \langle g_1, g_3|\, (g_3, g_4|.
\end{eqnarray}

\subsection{Symmetries of the reduced density matrix}
We now show that this model has a $G$ symmetry, by deriving its local action on the purification $|\Psi\rangle$ [for the results, see Eqs.~(\ref{supp_eq:lpdo_weak_strong},\ref{supp_eq:strong_L}).

The symmetry action is generated by,
\begin{equation}
\hat{U}_{g}=\begin{array}{c}
	\begin{tikzpicture}[scale=.4, baseline={([yshift=0ex]current bounding box.center)}, thick]
		\ATensorMPO{0, 1}{$A_g$}
        \ATensorMPO{2, 1}{$A_g$}
        \draw[shift={(0,0)},dotted] (2.8,1) -- (5,1);
        \ATensorMPO{6, 1}{$A_g$}
        \draw[shift={(0,2.8)}] (-1.2,-1.8) arc(90:270:0.2);
         \draw[shift={(0,0)}] (7.2,0.6) arc(-90:90:0.2);
		\draw [decorate,
    	decoration = {calligraphic brace,mirror}] (-0.5,-0.5) --  (6.8,-0.5);
		\draw (2.7,-1) node {\scriptsize $N$};  
	\end{tikzpicture}\end{array}, 
\end{equation}
and 
\begin{equation}
  \begin{array}{c}
\begin{tikzpicture}[scale=0.4, baseline={([yshift=-1.9ex]current bounding box.center)}, thick]
       \ATensorMPO{0, 0}{$A_g$}
  \end{tikzpicture}
  \end{array}=
  \begin{array}{c}
  \begin{tikzpicture}[scale=0.4, baseline={([yshift=-1.9ex]current bounding box.center)}, thick]
       \draw (-1, -1.5)--(-1, 1.5);
       \draw (1, -1.5)--(1, 1.5);
       \draw (-2.4, -1)--(-1, -1);
       \draw (2.4, -1)--(1, -1);
       \xTensor{0, 0}{$\alpha_g$}
       \wTensor{-1, 1}{$R_g$}
       \wTensor{1, 1}{$R_g$}
  \end{tikzpicture}
  \end{array}.\label{supp_eq:rd_spt_RGFP}
\end{equation}
Here, the trivalent junction corresponds to a Dirac-delta function, i.e., $\begin{array}{c} \begin{tikzpicture}[scale=0.4, baseline={([yshift=-0.4ex]current bounding box.center)}, thick]
		\draw (-0.5,1) -- (0,1);
		\draw (0, 0.5) -- (0,1.5);
        \draw (0,0.2)   node {\scriptsize $g_1$}; 
        \draw (0, 1.8)   node {\scriptsize $g_2$}; 
        \draw (-0.8, 1.1)   node {\scriptsize $g_3$}; 
	\end{tikzpicture}
	\end{array}\equiv\delta_{g_1, g_2, g_3}$. The tensor $\begin{tikzpicture}[scale=0.4, baseline={([yshift=-0.8ex]current bounding box.center)}, thick]
		\xTensor{0, 0}{$\alpha_g$}
        \draw (-1, -0.5)--(-1, 0.5);
        \draw (1, -0.5)--(1, 0.5);
	\end{tikzpicture}$ is diagonal, and built from a function $\alpha:G\times G\times G\rightarrow U(1)$, with the standard pentagon equation,
    \begin{equation}
\frac{\alpha(g_{12},g_3, g_4)\alpha(g_1,g_2,g_{34})}{\alpha(g_1,g_2, g_3)\alpha(g_1,g_{23}, g_4)\alpha(g_2,g_3, g_4)}=1. \label{supp_eq:pentagon_spt_rdm}
\end{equation}
Concretely,
\begin{equation}
\alpha_g\ |g_l, g_r\rangle\equiv \alpha[g_lg_r^{-1}, R_g (g_r), g]\ |g_l, g_r\rangle,\label{supp_eq:alpha_g}
\end{equation} 
where $R_g$ is the right group action, i.e., $R_g(g_1)=g_1g^{-1}$. 

The symmetry manifests as the push-through identity,
\begin{equation}\label{supp_eq:lpdo_weak_strong}
\begin{tikzpicture}[scale=.4, baseline={([yshift=-1.2ex]current bounding box.center)}, thick]
  \ATensorLPDO{0,0}{$L$}
\end{tikzpicture}
=
\begin{tikzpicture}[scale=.4, baseline={([yshift=-2.4ex]current bounding box.center)}, thick]
  \ATensorLPDO{0,0}{$L$}
  \wTensor{-0.6,0.9}{$w_g$}
  \wTensora{0.6,0.9}{$w^{\scalebox{0.5}{$a$}}_{\scalebox{0.5}{$g$}}$}
  \xTensor{2, 0}{$V^{\scalebox{0.5}{$\dagger$}}_{\scalebox{0.5}{$g$}}$}
  \xTensor{-2, 0}{$V_g$}
\end{tikzpicture}.
\end{equation}
Here, 
\begin{equation}\label{supp_eq:w_gsptPEPS}
w_g=
 \begin{array}{c}
  \begin{tikzpicture}[scale=0.4, baseline={([yshift=-1.9ex]current bounding box.center)}, thick]
       \draw (-1, -1)--(-1, 1);
       \draw (1, -1)--(1, 1);
       \xTensor{0, 0}{$\alpha_g$}
       \wTensor{-1, 0.5}{$R_g$}
       \wTensor{1, 0.5}{$R_g$}
  \end{tikzpicture}
  \end{array},\ w_g^a = w_g,\ \text{and}\ V_g=w_g.
\end{equation}
The derivation is achieved by proving the following graphical identity,
\begin{equation}\label{supp_eq:alpha_weaksym}
\begin{tikzpicture}[scale=.4, baseline={([yshift=-0.6ex]current bounding box.center)}, thick]
     \SPTbdTensorLPDOh{0, 0}{$g_1 g^{-1} $}{$\ g_2 g^{-1}$}{$g_4 g^{-1}$}{$g_3 g^{-1}$}
  \end{tikzpicture}=
  \begin{tikzpicture}[scale=.4, baseline={([yshift=-0.6ex]current bounding box.center)}, thick]
     \SPTbdTensorLPDO{0, 0}{$g_1 $}{$g_2$}{$g_4$}{$g_3$}

      \draw (-0.6, 2)--(0.6, 2);
     \alphaTensor{0, 2}{\scalebox{0.75}{$\alpha_g$}}

     \draw (-0.6, -2)--(0.6, -2);
     \alphaTensor{0,-2}{\scalebox{0.6}{$\alpha_g^\dagger$}}

     \draw (-2, -0.6)--(-2, 0.6);
     \alphaTensor{-2,0}{\scalebox{0.6}{$\alpha_g^\dagger$}}

     \draw (2, -0.6)--(2, 0.6);
     \alphaTensor{2,0}{\scalebox{0.6}{$\alpha_g$}}

  \end{tikzpicture},
\end{equation}
which indicates that $\alpha_g$ ($\alpha_g^\dagger$) shifts the PEPS tensor by $g^{-1}$, and thus reproduces Eq.~\eqref{supp_eq:lpdo_weak_strong}: applying $R_g^\dagger$ ($R_g$) to $|g_1, g_2)$ and $|g_2, g_4\rangle$ ($(g_3, g_4|$ and $\langle g_1, g_3|$) compensates the shift by $\alpha_g^\dagger$ ($\alpha_g$).
Concretely, we proceed in three steps: (i) We apply $\alpha_g^\dagger$ to the bottom-left triangle PEPS tensor [see the right-hand side term in Eq.~\eqref{supp_eq:alpha_weaksym}], whose main effect is to shift the index $g_1, g_2, g_3$ by $g^{-1}$, i.e.,  
\begin{equation}\label{supp_eq:tri_peps_shift}
\alpha(g_1g_2^{-1}, g_2g_4^{-1}, g_4)\rightarrow\alpha(g_1g_2^{-1}, g_2g_4^{-1}, g_4g^{-1}).
\end{equation}
Here, the indices $g_l, g_r$ in $\alpha_g$ [Eq.~\eqref{supp_eq:alpha_g}] are indicated by the edge arrow, pointing from $g_r$ to $g_l$.
We have used the following identity,

\begin{eqnarray}\label{supp_eq:alpha_bottom_left}
&&\alpha(g_1g_2^{-1},g_2g_4^{-1},g_4g^{-1})\nonumber\\
&=&\alpha(g_1g_2^{-1},g_2g_4^{-1},g_4)\nonumber\\
&&\times\frac{1}{\alpha(g_2g_4^{-1}, g_4g^{-1}, g)\alpha(g_1g_2^{-1}, g_2g^{-1}, g)},\nonumber\\
&&\times \alpha(g_1g_4^{-1},g_4g^{-1}, g),
\end{eqnarray}
obtained via the standard pentagon equation of $\alpha$ for the group elements $(g_1g_2^{-1}, g_2g_4^{-1}, g_4g^{-1}, g)$. The factor on the third line, $\frac{1}{\alpha(g_2g_4^{-1}, g_4g^{-1}, g)\alpha(g_1g_2^{-1}, g_2g^{-1}, g)}$, comes from the action of $\alpha_g^\dagger$ on the left and the bottom edges. The final factor $\alpha(g_1g_4^{-1},g_4g^{-1}, g)$ instead comes from the action of $\alpha_g$ on $|g_1, g_4)$, and is canceled by the contributions from the top-right triangle in Eq.~\eqref{supp_eq:alpha_weaksym} (see below). (ii) We then apply $\alpha_g$ to the remaining top-right triangle. Again, its main effect is to shift the group elements by $g^{-1}$, as seen from the following identity implied by the pentagon equation,
\begin{eqnarray}\label{supp_eq:alpha_top_right}
&&\frac{1}{\alpha(g_1g_3^{-1},g_3g_4^{-1},g_4g^{-1})}\nonumber\\
&=&\frac{1}{\alpha(g_1g_3^{-1},g_3g_4^{-1},g_4)}\nonumber\\
&&\times \alpha(g_3g_4^{-1}, g_4g^{-1}, g)\alpha(g_1g_3^{-1}, g_3g^{-1}, g)\nonumber\\
&&\times \frac{1}{\alpha(g_1g_4^{-1},g_4g^{-1}, g)},
\end{eqnarray}
where the factor on the third line (the final) comes from $\alpha_g$ ($\alpha_g^\dagger$) on the top and the right edges ($|g_1, g_4)$). (iii) Combining Eq.~\eqref{supp_eq:alpha_bottom_left} with Eq.~\eqref{supp_eq:alpha_top_right}, we verify the identity in Eq.~\eqref{supp_eq:alpha_weaksym}. Explicitly, this follows from taking their product,
\begin{eqnarray}
&&\frac{\alpha(g_1g_2^{-1},g_2g_4^{-1},g_4g^{-1})}{\alpha(g_1g_3^{-1},g_3g_4^{-1},g_4g^{-1})}\nonumber\\
&=&\frac{\alpha(g_1g_2^{-1},g_2g_4^{-1},g_4)}{\alpha(g_1g_3^{-1},g_3g_4^{-1},g_4)}\nonumber\\
&&\times\frac{\alpha(g_3g_4^{-1}, g_4g^{-1}, g)\alpha(g_1g_3^{-1}, g_3g^{-1}, g)}{\alpha(g_2g_4^{-1}, g_4g^{-1}, g)\alpha(g_1g_2^{-1}, g_2g^{-1}, g)},\nonumber\\
\end{eqnarray}
which reproduces Eq.~\eqref{supp_eq:alpha_weaksym}.

Additionally, we find the following push-through identity,
\begin{equation}\label{supp_eq:strong_L}
\begin{tikzpicture}[scale=.4, baseline={([yshift=-2.4ex]current bounding box.center)}, thick]
  \ATensorLPDO{0,0}{$L$}
  \wTensor{-0.6,0.9}{$\scalebox{0.8}{$d$}_{\scalebox{0.5}{${1,2}$}}$}
\end{tikzpicture}
=
\begin{tikzpicture}[scale=.4, baseline={([yshift=-1.2ex]current bounding box.center)}, thick]
  \ATensorLPDO{0,0}{$L$}
  \xTensor{-2, 0}{$\scalebox{0.7}{$\Omega$}^{\scalebox{0.5}{$\dagger$}}_{\scalebox{0.4}{$1,2$}}$}
 \xTensor{2, 0}{$\scalebox{0.7}{$\Omega$}_{\scalebox{0.4}{$1,2$}}$}
\end{tikzpicture}.
\end{equation}
Here, $d_{1,2}\equiv w_{g_1}w_{g_2}w_{g_1g_2}^\dagger$, i.e., 
\begin{eqnarray}
&&d_{1,2}|g_l, g_r\rangle\nonumber\\
&=& \frac{\alpha(g_lg_r^{-1}, g_r, g_1)\ \alpha(g_lg_r^{-1}, g_rg_1, g_2)}{ \alpha(g_lg_r^{-1}, g_r , g_1g_2)}|g_l, g_r\rangle\nonumber\\
{}&=& \alpha(g_l, g_1, g_2)\times \alpha(g_r, g_1, g_2)^* |g_l, g_r\rangle,\label{supp_eq:d_12_spt_rdm}
\end{eqnarray}
where, in the last line, we have used the pentagon equation [Eq.~\eqref{supp_eq:pentagon_spt_rdm}] for $(g_l g_r^{-1},g_r, g_1, g_2)$. It is interesting to note that $d_{1,2}$ is diagonal in the two-site product basis $|g_1\rangle\otimes |g_2\rangle$.  Also, $\Omega_{1,2}$, defined in Eq.~\eqref{supp_eq:strong_L}, is diagonal, i.e.,
\begin{equation}
    \Omega_{1,2}|g)=e^{i\theta_{1,2}(g)}|g),\ \text{with}\ e^{i\theta_{1,2}(g)}\equiv\alpha(g, g_1, g_2)^*,
\end{equation}
and the round bracket $|g)$ is used for the basis of the virtual space. This leads to the following identity from a crossed module extension~\cite{else2014prb},
\begin{equation}
\Omega_{1,2}\Omega_{12,3}=\alpha(g_1, g_2, g_3)\ {}^1\Omega_{2,3}\Omega_{1,23},
\end{equation}
with 
\begin{equation}
{}^1\Omega_{2,3}\equiv V_{g_1} \Omega_{2,3} V_{g_1}^\dagger,
\end{equation}
derived from the two equations below,
\begin{equation}
\Omega_{1,2}\Omega_{12,3}|g)=\alpha(g, g_1, g_2)^*\alpha(g, g_1g_2, g_3)^*|g),
\end{equation}
and 
\begin{equation}
{}^1\Omega_{2,3}\Omega_{1,23}|g)=\alpha(gg_1, g_2, g_3)^*\alpha(g, g_1, g_2g_3)^*|g).
\end{equation}

\subsection{Derivation of the reduced density matrix \label{supp_sec:rdm_lpdo}}
We now derive Eq.~\eqref{supp_eq:peps_lpdo}. To build an intuition, we first present the method for one-dimensional short-range correlated states, and then extend it to two dimensions. 

To this end, consider a one-dimensional translation-invariant MPS at its renormalization-group (RG) fixed point~\cite{cirac2017aop}, i.e., 
\begin{equation}
|\psi\rangle=
	\begin{tikzpicture}[scale=.4, baseline={([yshift=0.3ex]current bounding box.center)}, thick]
        \draw[shift={(0,0)},dotted] (2,0) -- (5,0);
    
		\ATensorMPS{0, 0}{$A$}
        \ATensorMPS{1.8, 0}{$A$}
        
        \ATensorMPS{5.7, 0}{$A$}
        
        \draw[shift={(0,0.8)}] (-1.2,-0.8) arc(90:270:0.2);
         \draw[shift={(-1,0)}] (7.8,-0.4) arc(-90:90:0.2);
		\draw [decorate,
    	decoration = {calligraphic brace,mirror}] (-1,-0.8) --  (6.8,-0.8);
		\draw (2.7,-1.5) node {\scriptsize $N$};  
        
	\end{tikzpicture}\ ,
\end{equation}
with the transfer matrix $
T\equiv 
    \begin{tikzpicture}[scale=.3, baseline={([yshift=-0.4ex]current bounding box.center)}, thick]
    \ATensorMPS{0, -1}{$A$}
    \AdagTensorMPS{0, 1}{$A^*$}
    \end{tikzpicture}$.
At the RG fixed point, $TT=T$, so $T$ has a single non-zero eigenvalue,
\begin{equation}
\begin{array}{c}
  \begin{tikzpicture}[scale=.4, baseline={([yshift=-2.4ex]current bounding box.center)}, thick]
    \ATensorMPS{0,-1}{$A$} 
    \AdagTensorMPS{0, 1}{$A^*$} ;
  \end{tikzpicture}\end{array}=
	\begin{array}{c}
		\begin{tikzpicture}[scale=.4, baseline={([yshift=-1.4ex]current bounding box.center)}, thick]
		\draw[rounded corners] (0,2.4)--(0.5,2.4)--(0.5,0.4)--(0,0.4);
		\wTensor{0.5,0.9}{$\sigma_{R}$}
		\draw[rounded corners] (2.2,2.4)--(1.7,2.4)--(1.7,0.4)--(2.2,0.4);
		\end{tikzpicture}
	\end{array}\ \text{with}\ 
    	\begin{array}{c}
		\begin{tikzpicture}[scale=.4, baseline={([yshift=-1.4ex]current bounding box.center)}, thick]
		\draw[rounded corners] (1,2.4)--(0.5,2.4)--(0.5,0.4)--(1,0.4);
		
		\draw[rounded corners] (1.2,2.4)--(1.7,2.4)--(1.7,0.4)--(1.2,0.4);

        \wTensor{1.7,0.9}{$\sigma_{R}$}
        \draw (1, 2.4)--(1.2, 2.4);
        \draw (1, 0.4)--(1.2, 0.4);
		\end{tikzpicture}=1
        \end{array},
\end{equation}
where the left eigenvector is $\begin{tikzpicture}[scale=.3, baseline={([yshift=-0.6ex]current bounding box.center)}, thick]
		\draw[rounded corners] (2.2,2.4)--(1.7,2.4)--(1.7,0.4)--(2.2,0.4);
		\end{tikzpicture}$, fixed by a gauge choice~\cite{cirac2021rmp}.
        
The reduced density matrix is then given by
\begin{equation}
    \begin{tikzpicture}[scale=.4, baseline={([yshift=-0.8ex]current bounding box.center)}, thick]

   \draw (0, -1.2)--(0, 1.2);

		\wTensor{0,-0.5}{$\rho_{\scalebox{0.5}{RM}}$}
    \end{tikzpicture}
    =
    \begin{tikzpicture}[scale=.4, baseline={([yshift=-1.2]current bounding box.center)}, thick]
    
    \ATensorMPS{0, -1}{$A^*$}
    \AdagTensorMPS{0, 1}{$A$}
    \draw (1.2, 1)--(1.2, -1);
    
    \draw[red] (0, -0.4)--(0, 0.4);
    \draw[red] (0.6, 1)--(1.2, 1);
    \draw[red] (0.6, -1)--(1.2, -1);
     \draw[red] (1.2, -1)--(1.2, 1);

     \draw (-1.2, 1)--(-1.2, 1.6);
     \draw (-1.2, -1)--(-1.2, -1.6);
     
    \end{tikzpicture}\ ,\label{supp_eq:rho_rdm_1d}
\end{equation}
where an isometry that maps the physical space to the virtual space is applied, and the red line denotes the ancilla in the LPDO representation. 
Specifically, a reduced density matrix is defined by tracing out one half of the system,
\begin{equation}
\begin{tikzpicture}[scale=.4, baseline={([yshift=-1ex]current bounding box.center)}, thick]
    \draw[dotted](2, 1)--(3.5, 1);
    \draw[dotted](2, -1)--(3.5, -1);
    \draw[dotted](-2, 1)--(-3.5, 1);
    \draw[dotted](-2, -1)--(-3.5, -1);
    \ATensorMPS{1, -1}{$A^*$}
    \AdagTensorMPS{1, 1}{$A$}
     \AdagTensorMPS{-1, -1}{$A^*$}
    \ATensorMPS{-1, 1}{$A$}

    \draw[red, dashed](0,-2)--(0, 2);
    \draw[red, dashed](3.3,-2)--(3.3, 2);

    \draw[shift={(0,1.8)}] (-3.7,-0.8) arc(90:270:0.2);
     \draw[shift={(0,-1)}] (-3.7,0) arc(90:270:0.2);
    \draw[shift={(0,0.6)}] (3.5,0) arc(-90:90:0.2);
    \draw[shift={(0,-1.4)}] (3.5,0) arc(-90:90:0.2);
         
    \end{tikzpicture}=
    \begin{tikzpicture}[scale=.4, baseline={([yshift=-1ex]current bounding box.center)}, thick]
    \draw[dotted](3, 1)--(4.5, 1);
    \draw[dotted](3, -1)--(4.5, -1);
    \draw[dotted](-3, 1)--(-4.5, 1);
    \draw[dotted](-3, -1)--(-4.5, -1);
    
    \AdagTensorMPS{2, -1}{$A^*$}
    \ATensorMPS{2, 1}{$A$}
    
    \ATensorMPS{0, -1}{$A^*$}
    \AdagTensorMPS{0, 1}{$A$}
    
     \AdagTensorMPS{-2, -1}{$A^*$}
    \ATensorMPS{-2, 1}{$A$}

    \draw[shift={(-1,1.8)}] (-3.7,-0.8) arc(90:270:0.2);
     \draw[shift={(-1,-1)}] (-3.7,0) arc(90:270:0.2);
    \draw[shift={(1,0.6)}] (3.5,0) arc(-90:90:0.2);
    \draw[shift={(1,-1.4)}] (3.5,0) arc(-90:90:0.2);

    \draw[red, dashed](-1,-2)--(-1, 2);
    \draw[red, dashed](1,-2)--(1, 2);
    \end{tikzpicture},\label{supp_eq:rm_spectrum}
\end{equation}
where the red dashed line marks the bipartition, and the equality follows from the RG fixed-point condition. To obtain the reduced density matrix, we observe that the spectra of $M_1 M_2$ and $M_2 M_1$ for two matrices $M_1, M_2$ are the same, so viewing the right-hand side as a product of two matrices yields the following reduced density matrix, up to an isometry,
\begin{equation}
\begin{tikzpicture}[scale=.4, baseline={([yshift=-1ex]current bounding box.center)}, thick]
    \draw[dotted](3, 1)--(4.5, 1);
    \draw[dotted](3, -1)--(4.5, -1);
    \draw[dotted](-3, 1)--(-4.5, 1);
    \draw[dotted](-3, -1)--(-4.5, -1);
    
    \AdagTensorMPS{2, -1}{$A^*$}
    \ATensorMPS{2, 1}{$A$}

     \AdagTensorMPS{-2, -1}{$A^*$}
    \ATensorMPS{-2, 1}{$A$}

    \draw[shift={(-1,1.8)}] (-3.7,-0.8) arc(90:270:0.2);
     \draw[shift={(-1,-1)}] (-3.7,0) arc(90:270:0.2);
    \draw[shift={(1,0.6)}] (3.5,0) arc(-90:90:0.2);
    \draw[shift={(1,-1.4)}] (3.5,0) arc(-90:90:0.2);

    \draw[dashed, red](-1, -0.8)--(1, -0.8);

    \draw[rounded corners,shift={(-1,0)}] (0,1)--(0.5,1)--(0.5,-1)--(0,-1);
		\wTensor{-0.5,-0.5}{$\sigma_{R}^{\scalebox{0.5}{T}}$}
		\draw[rounded corners,shift={(-1.2,0)}] (2.2,1)--(1.7,1)--(1.7,-1)--(2.2,-1);

    \end{tikzpicture}
    \implies 
    \begin{tikzpicture}[scale=.4, baseline={([yshift=-1ex]current bounding box.center)}, thick]

    \draw[rounded corners,shift={(-1,0)}] (0,1)--(0.5,1)--(0.5,-1)--(0,-1);
		\wTensor{-0.5,-0.5}{$\sigma_{R}$}
		\draw[rounded corners,shift={(-0.7,0)}] (2.2,1)--(1.7,1)--(1.7,-1)--(2.2,-1);
        \wTensor{1,-0.5}{$\sigma_{R}^{\scalebox{0.5}{$T$}}$}

    \end{tikzpicture},
\end{equation}
representing the effective edge density matrix at both ends of the bipartition. This reproduces the reduced density matrix for a single edge in Eq.~\eqref{supp_eq:rho_rdm_1d}.

For two-dimensional systems, we repeat the 1D derivation, by replacing $\begin{tikzpicture}[scale=.4, baseline={([yshift=-1.4ex]current bounding box.center)}, thick] \ATensorMPS{0, 0}{$A$}\end{tikzpicture}$ in Eq.~\eqref{supp_eq:rho_rdm_1d} with a one-dimensional MPS along the transverse direction. Namely, the RG fixed point PEPS tensor for a two-dimensional SPT is~\cite{willamson2016prb}
\begin{equation}
\begin{tikzpicture}[scale=.4, baseline={([yshift=-0.6ex]current bounding box.center)}, thick]
     \SPTbulkTensorPEPS{0, 0}{$g_1 $}{$g_2$}{$g_4$}{$g_3$}
  \end{tikzpicture}
  =\frac{\alpha(g_1g_2^{-1}, g_2g_4^{-1}, g_4)}{\alpha(g_1g_3^{-1}, g_3g_4^{-1}, g_4)},
\end{equation}
where the thick line denotes the physical index, and the trivalent node on the black lines enforces a Dirac delta. The corresponding LPDO [Eq.~\eqref{supp_eq:peps_lpdo}] is,
\begin{equation}
\begin{tikzpicture}[scale=.4, baseline={([yshift=-0.6ex]current bounding box.center)}, thick]
     \SPTbdTensorLPDO{0, 0}{$g_1 $}{$g_2$}{$g_4$}{$g_3$}
  \end{tikzpicture}
=\frac{\alpha(g_1g_2^{-1}, g_2g_4^{-1}, g_4)}{\alpha(g_1g_3^{-1}, g_3g_4^{-1}, g_4)},
\end{equation}
where the external physical index (dark thick line) is omitted, as it cancels in the sum imposed by the Dirac delta at the trivalent junction.

\section{More on the large gauge transformations in tensor networks \label{supp_sec:flux_tn}}
Here, we provide further details on the large-gauge-transformation operators for the bosonic models studied in the main text. Specifically, we (i) show that any cocycle $\alpha$ can be brought to a standard form by a shallow unitary transformation, and (ii) briefly review the 3-cocycle $\alpha$ for Abelian symmetries.

\subsection{Bringing $\alpha$ to standard form via a shallow unitary \label{supp_sec:standard_unitary}}
We now show that $\alpha$ can be brought to a standard form by a shallow-depth unitary. Since each equivalence class admits such a representative, it suffices to prove that ${\hat{U}_g}$ are unitarily equivalent under changes of $\alpha$ within a given class, for $g\in G$ with $G$ a general (not necessarily Abelian) symmetry group. Specifically, the equivalence relation for $\alpha$ is given by,

\begin{equation}
\alpha^\prime\left(g_{1},\ g_{2},\ g_{3}\right)\simeq \alpha\left(g_{1},\ g_{2},\ g_{3}\right),\  g_{1,2,3}\in G,
\end{equation}
where 
\begin{equation}
\alpha^\prime\left(g_{1},\ g_{2},\ g_{3}\right)\equiv \alpha\left(g_{1},\ g_{2},\ g_{3}\right)\times\left(\delta\gamma\right)\left(g_{1},\ g_{2},\ g_{3}\right),
\end{equation}
and 
\begin{equation}
\left(\delta\gamma\right)\left(g_{1},\ g_{2},\ g_{3}\right)\equiv\frac{\gamma\left(g_{2},\ g_{3}\right)\gamma\left(g_{1},\ g_{2}g_{3}\right)}{\gamma\left(g_{1},\ g_{2}\right)\gamma\left(g_{1}g_{2},\ g_{3}\right)},
\end{equation}
with $\gamma: G\times G\rightarrow U(1)$. Then, for $\left\{\hat{U}_g^\prime\right\}$ and $\left\{\hat{U}_g\right\}$ constructed from $\alpha^\prime$ and $\alpha$, respectively, the unitary connecting them is 
\begin{equation}\label{supp_eq:U_g_trans}
\hat{U}_g^\prime  = \hat{\mathcal{U}}^\dagger \hat{U}_g\hat{\mathcal{U}},
\end{equation}
where 
\begin{equation}
\hat{\mathcal{U}}=
\begin{array}{c}
  \begin{tikzpicture}[scale=0.4, baseline={([yshift=-1.9ex]current bounding box.center)}, thick]
       \draw (-1, -1)--(-1, 1);
       \draw (1, -1)--(1, 1);
       \draw (-1.5, 0)--(0, 0);
       \xTensor{0, 0}{\scalebox{0.9}{$m_\gamma$}}

        \xTensor{2, 0}{\scalebox{0.9}{$m_\gamma$}}

         \draw (3, -1)--(3, 1);

         \draw[dotted](3, 0)--(6, 0);
         \draw(3, 0)--(3.5, 0);
         \draw(5.5, 0)--(6, 0);

         \draw (6, -1)--(6, 1);
          \xTensor{7, 0}{\scalebox{0.9}{$m_\gamma$}}
           \draw (8, -1)--(8, 1);

           \xTensor{9, 0}{\scalebox{0.9}{$m_\gamma$}}
          
           \draw[shift={(0,0)}] (-1.5,0) arc(90:270:0.2);
           
         \draw[shift={(0,-1)}] (10,0.6) arc(-90:90:0.2);
		
  \end{tikzpicture}
  \end{array},
\end{equation}
is a unitary with shallow depth, because
$m_\gamma$ acts as $m_\gamma|g_l , g_r\rangle=\gamma\left(g_{l}g_{r}^{-1},\ g_{r}\right)|g_{l},\ g_{r}\rangle$ and its action on different sites commutes. The proof of Eq.~\eqref{supp_eq:U_g_trans} is then based on the following identity
\begin{eqnarray}
w_{g}^{\prime}|g_{l},\ g_{r}\rangle&=&\left[\frac{1}{\gamma\left(g_{l}g^{-1},\ g\right)}\gamma\left(g_{r}g^{-1},\ g\right)\right]\nonumber\\
&&\times\left(m_{\gamma}^{\dagger}w_{g}m_{\gamma}\right)|g_{l},\ g_{r}\rangle,
\end{eqnarray}
which relates the local building blocks $w_g^\prime$ of $\hat U_g^\prime$ and $w_g$ of $\hat U_g$.
Here $m_\gamma$ is a two-site tensor, while $\gamma(g_l g^{-1},g)^{-1}$ and $\gamma(g_r g^{-1},g)$ act as single-site factors. Consequently, these single-site factors cancel in the full matrix-product operator $\hat{\mathcal U}$, yielding Eq.~\eqref{supp_eq:U_g_trans}.

\subsection{A brief summary of $3$-cocycle $\alpha$ for Abelian symmetries \label{supp_sec:3cocycle}}
Here we briefly summarize useful results for $\alpha(g_1,g_2,g_3)$ \cite{propitius1995thesis} with $g_{1,2,3}\in G$. We parameterize $\hat U_{g_1}\equiv e^{i\boldsymbol a\cdot\hat{\boldsymbol Q}}$, and similarly $\hat U_{g_2}\equiv e^{i\boldsymbol b\cdot\hat{\boldsymbol Q}}$ and $\hat U_{g_3}\equiv e^{i\boldsymbol c\cdot\hat{\boldsymbol Q}}$, as in the main text.

The 3-cocycle $\alpha$ [more precisely, $H^{3}(G,U(1))$ for finite Abelian $G$] decomposes into three families, called Type I, II, and III~\cite{propitius1995thesis}. This is because any finite Abelian group $G$ can be written as $\mathbb{Z}_{\mathcal{N}_1}\times \mathbb{Z}_{\mathcal{N}_2}\dots$, up to an isomorphism. In turn, its third cohomology group factorizes as
\begin{eqnarray}
H^{3}(G, U(1))&=&\overset{\text{Type I}}{\underline{\prod_{I}\mathbb{Z}_{\mathcal{N}_{I}}}}\ \ \overset{\text{Type II}}{\underline{\prod_{I<J}\mathbb{Z}_{\gcd(\mathcal{N}_{I},\mathcal{N}_{J})}}}\ \ \nonumber\\
&&\times \overset{\text{Type III}}{\underline{\prod_{I<J<K}\mathbb{Z}_{\gcd(\mathcal{N}_{I},\mathcal{N}_{J},\mathcal{N}_{K})}}}
\end{eqnarray}
Here $\gcd(\cdot)$ denotes the greatest common divisor, and $I, J,K$ label the cyclic groups. The standard form of $\alpha$ for each piece is given by
\begin{itemize}
    \item Type I: $\alpha(\boldsymbol{a}, \boldsymbol{b}, \boldsymbol{c})=e^{\frac{2}{4\pi}i\sum_{I}p_{I}\,a_{I}\,\Delta_{I}\left(\boldsymbol{b},\boldsymbol{c}\right)}$, where the components $a_{I},b_{I},c_{I}\in\frac{2\pi}{\mathcal{N}_I}\{0,1,\dots,\mathcal{N}_{I}-1\}$ label the $\mathbb{Z}_{\mathcal{N}_{I}}$ factors of
$\boldsymbol{a},\boldsymbol{b},\boldsymbol{c}$. $\Delta_{I}(\boldsymbol{b},\boldsymbol{c})=\left(b_{I}+c_{I}-[b_{I}+c_{I}]_{2\pi}\right)$ with $[\cdot]_{2\pi}$ denotes modulo of $2\pi$. Additionally, $p_I\in \mathbb{Z}_{\mathcal{N}_I}$ because $\alpha$ is invariant under $p_i\rightarrow p_I+\mathcal{N}_I$ since $\Delta_I\in 2\pi \mathbb{Z}$.
\item Type II:  $\alpha(\boldsymbol{a}, \boldsymbol{b}, \boldsymbol{c})=e^{\frac{2}{4\pi}i\sum_{I<J}q_{IJ}\,a_{I}\,\Delta_{J}\left(\boldsymbol{b},\boldsymbol{c}\right)}$, where $I, J$ label distinct cyclic groups. The coefficient $q_{IJ}\in \mathbb{Z}_{\gcd(\mathcal{N}_I, \mathcal{N}_J)}$ is defined modulo $\gcd(\mathcal{N}_I, \mathcal{N}_J)$, since $\alpha$ is invariant under $q_{IJ}\rightarrow q_{IJ}+\gcd(\mathcal{N}_I, \mathcal{N}_J)$.
\item Type III: $\alpha(\boldsymbol{a}, \boldsymbol{b}, \boldsymbol{c})=e^{\frac{2}{4\pi}i\sum_{I<J<K}\frac{1}{\left(2\pi\right)^{2}}\frac{l_{IJK}\mathcal{N}_{I}\mathcal{N}_{J}\mathcal{N}_{K}}{\gcd(\mathcal{N}_{I},\mathcal{N}_{J},\mathcal{N}_{K})}\,a_{I}b_{J}c_{K}}$, with $l_{IJK}\in \mathbb{Z}_{\gcd(\mathcal{N}_I, \mathcal{N}_J, \mathcal{N}_K)}$.

\end{itemize}
Hence, these three sets of integers $p_I, q_{IJ}$, and $l_{IJK}$ parametrize the Type I/II/III components of $\alpha$ for the group $G$.